\documentclass[trackchanges,twocolumn]{aastex701}
\usepackage{amsmath} 
\usepackage{url}
\usepackage[normalem]{ulem}
\usepackage{cancel}

\def\aap{Astronomy and Astrophysics}

\def\apjs{ApJS}
\def\apjl{ApJL}

\graphicspath{ {./figs/} }

\begin{document}

\title{Faraday Complexity and Depolarisation in a High-Rotation-Measure Radio Galaxy from the Spectra and Polarisation In Cutouts of Extragalactic Sources (SPICE-RACS) DR2}

\author[]{Debajyoti Mondal}
\affiliation{Indian Institute of Engineering Science and Technology, Botanical Garden Area, Howrah, West Bengal 711103, India}
\email[show]{djmondal596@gmail.com}

\author[orcid=0000-0002-8566-9955]{Abhik Ghosh}
\affiliation{Department of Physics, Banwarilal Bhalotia College, Asansol, West Bengal, Pin: 713303, India}
\email[show]{abhik.physicist@gmail.com}  

\author[orcid=0009-0008-3561-9926]{Dipanjan Banerjee}
\affiliation{Department of Physics, Banwarilal Bhalotia College, Asansol, West Bengal, Pin: 713303, India}
\email[]{dipanjanbanerjee20@gmail.com} 

\begin{abstract}
We present a broadband spectro-polarimetric analysis of the extragalactic radio source \texttt{RACS\_0900-28\_7036} using SPICE-RACS DR2 observations with the Australian Square Kilometre Array Pathfinder (ASKAP). The source was selected for its large rotation measure (${\rm RM}=345.7\pm0.2~{\rm rad~m^{-2}}$), substantial excess relative to the local foreground ($\Delta {\rm RM}\approx171~{\rm rad~m^{-2}}$), and strong evidence of Faraday complexity ($\sigma_{\rm add}/\delta\sigma_{\rm add}\approx8.6$). Observations span 803--1083~MHz in 36 spectral channels, enabling detailed characterization of Faraday rotation and wavelength-dependent depolarization. One-dimensional QU-fitting and Bayesian model selection identify a multi-component model comprising one Burn-slab component and two external Faraday dispersion components (1 Slab + 2 EFD) as the preferred description. The dominant astrophysical component exhibits ${\rm RM}\approx345.5~{\rm rad~m^{-2}}$
with modest Faraday dispersion
($\sigma_{\rm RM}\approx3~{\rm rad~m^{-2}}$),
consistent with the Galactic foreground rotation measure at the source
position (${\rm RM}_{\rm Gal}=331.9\pm33.1~{\rm rad~m^{-2}}$).
A secondary broader component at
${\rm RM}\approx131.5~{\rm rad~m^{-2}}$
shows strong depolarization
($\sigma_{\rm RM}\approx19.5~{\rm rad~m^{-2}}$),
indicating an additional turbulent Faraday-active medium along the line of sight. The fractional polarization spectrum and $q$--$u$ plane evolution further confirm multiple Faraday-active regions along the line of sight. These results demonstrate that ASKAP broadband spectropolarimetry can resolve complex Faraday structures and probe turbulent magnetized environments, providing a framework for systematic depolarization studies across the full SPICE-RACS catalog and enabling statistical investigations of Faraday complexity in diverse extragalactic radio sources.
\end{abstract}


\section{Introduction}
\label{sec:intro}
Magnetic fields are a fundamental component of galaxies, galaxy groups, clusters, and the large-scale structure of the Universe. Radio polarimetric observations provide one of the most powerful tools for studying these fields by probing the magneto-ionic medium through the Faraday effect (e.g. \citealt{Taylor_2009, Mao_2010, Harvey_2011}). Large samples of extragalactic polarized radio sources have enabled the construction of increasingly dense rotation measure (RM) grids \citep{Loi25}, which can be used to investigate magnetic fields in the Milky Way, nearby galaxies, galaxy clusters, and the cosmic web \citep{Gaensler_2004, Dickey_2022, Livingston_2022, Anderson_2021, Carretti_2022, Shimwell_2022, Shimwell_2026}. In addition to serving as background probes, the polarization properties of radio galaxies themselves provide valuable information about the morphology, environment, and evolutionary state of active galactic nuclei (AGN) and their surrounding magnetised plasma \citep{OSullivan_2012, Pasetto_2016}.

When linearly polarized synchrotron radiation propagates through a magneto-ionic medium, the plane of polarization rotates by an amount proportional to the square of the observing wavelength. This phenomenon, known as Faraday rotation, depends on the integrated product of the thermal electron density and the line-of-sight magnetic field. At the same time, variations in Faraday rotation occurring within or across the emitting region can reduce the observed polarized intensity through a range of depolarization processes \citep{Burn_1966, Sokoloff_1998, Rossetti_2008, Pasetto_2018}. These effects encode information on the structure of the thermal plasma and magnetic field, including the relative contributions of ordered and turbulent components. Consequently, measurements of Faraday rotation and depolarization provide a unique diagnostic of the magneto-ionic environments both internal and external to radio galaxies \citep{Strom_1973}.

The advent of broadband radio spectro-polarimetry has transformed our ability to study such environments. By densely sampling polarization as a function of wavelength squared, modern surveys can distinguish between multiple Faraday-rotating components, identify Faraday-thick structures, and constrain the physical mechanisms responsible for depolarization \citep{Farnsworth_2011, Anderson_2015, OSullivan_2017, Pasetto_2018}. In particular, direct modelling of the complex polarization spectra through \emph{QU-fitting} has been shown to be highly sensitive to Faraday complexity and often recovers multiple components that are difficult to identify using RM synthesis alone \citep{Farnsworth_2011, OSullivan_2012, Sun_2015, OSullivan_2017, Paul_2026}.

Recent years have seen the emergence of large-area polarization surveys operating at both metre and centimetre wavelengths. Low-frequency surveys such as the \emph{LOFAR Two-Metre Sky Survey} (LoTSS; \citealt{Shimwell_2022, Oei_2023}) provide exceptional Faraday-depth resolution owing to their extensive $\lambda^{2}$ coverage, making them highly sensitive to Faraday-thin structures and weakly depolarized components. However, depolarization increases rapidly with wavelength, and broad Faraday-dispersive structures can become strongly suppressed at metre wavelengths. For example, in the case of external Faraday dispersion the fractional polarization follows
\[|p(\lambda^2)| \propto \exp(-2\sigma_{\rm RM}^{2}\lambda^{4}),\] 
such that even moderate RM dispersions can produce severe depolarization at low frequencies \citep{Burn_1966}. Consequently, low-frequency surveys are often insensitive to highly turbulent magneto-ionic media and Faraday-thick structures \citep{Gaensler_2025}.

Broadband GHz-frequency surveys provide an important complement. The \emph{Spectra and Polarisation In Cutouts of Extragalactic Sources} survey (SPICE-RACS; \citealt{Thomson_2023, Thomson_2026}), based on observations from the \emph{Australian Square Kilometre Array Pathfinder} (ASKAP; \citealt{Hotan_2021, McConnell_2020}), exploits the wide instantaneous bandwidth of ASKAP to obtain spectro-polarimetric measurements over approximately $799.5$--$1087.5$~MHz across a large fraction of the southern sky. Operating at higher frequencies than LoTSS, ASKAP remains sensitive to polarized emission from sources exhibiting substantial Faraday dispersion and complex Faraday-thick components that would be heavily depolarized at metre wavelengths. As a result, SPICE-RACS is particularly well suited to identifying radio galaxies embedded within turbulent or magnetically rich environments and to characterising Faraday complexity on scales inaccessible to low-frequency surveys alone.

The second data release of SPICE-RACS (SPICE-RACS DR2) provides polarization spectra, Faraday depth measurements, and derived polarization properties for a large population of extragalactic radio sources. This dataset offers an unprecedented opportunity to investigate the magneto-ionic environments of radio galaxies across a wide range of Faraday complexities. In particular, sources exhibiting unusually large rotation measures are of special interest because they are likely to trace dense thermal environments, strong magnetic fields, or multiple Faraday-active regions along the line of sight \citep{Anderson_2015, Anderson_2016, Anderson_2021}.

In this paper we investigate the Faraday complexity and depolarization properties of a highly polarized, high-RM radio galaxy selected from the SPICE-RACS DR2 catalogue. Using broadband ASKAP spectro-polarimetric data, we model the fractional Stokes parameters as functions of wavelength squared using a suite of physically motivated depolarization models. Our aim is to determine the minimum Faraday structure required to reproduce the observed polarization behaviour and to assess the extent to which unresolved polarization spectra can constrain the underlying magneto-ionic environment. This study serves as a pilot investigation for future systematic analyses of larger SPICE-RACS samples and provides insight into the physical origin of complex Faraday structures in powerful radio galaxies.

The remainder of this paper is organized as follows. Section~\ref{sec:target} presents the properties of the selected SPICE-RACS DR2 source. Section~\ref{sec:polFar} introduces the fundamental observables of radio polarization and Faraday rotation, together with the depolarization models employed in this work. Section~\ref{sec:methods} outlines the analysis methodology. Section~\ref{sec:results} presents the results of the polarization modelling, while Section~\ref{sec:discussion} discusses the implications for the source environment and prospects for future studies.

\section{Target Selection from SPICE-RACS DR2}
\label{sec:target}
The SPICE-RACS DR2 survey from the Rapid ASKAP Continuum Survey \citep{Thomson_2026} provides broadband spectropolarimetric observations of extragalactic radio sources with ASKAP, covering roughly $3.5\pi$ sr over $799.5$--$1087.5$~MHz. The survey includes $72$ or $36$ spectral channels at angular resolutions of $11.8$--$75.9$~arcsec, yielding a $\lambda^2$ coverage of $0.076$--$0.14~\mathrm{m^2}$ and a Faraday depth resolution of $\delta \phi \approx 63~\mathrm{rad~m^{-2}}$. This makes SPICE-RACS DR2 particularly suited for identifying sources with Faraday complexity and significant depolarisation. Typical Stokes $Q/U$ sensitivities reach $\sim 40~\mu\mathrm{Jy~beam^{-1}}$, enabling robust measurements of polarisation structure and rotation measures across thousands of radio galaxies \citep{Gaensler_2025}.

To select a promising candidate for detailed depolarisation studies, we developed a pipeline using the SPICE-RACS DR2 polarisation catalog\footnote{\url{https://data.csiro.au/collection/csiro:64891}} \citep{Thomson_2025}, focusing on extragalactic sources with significant Faraday rotation, evidence of Faraday complexity, and sufficiently high polarised signal-to-noise. Quality-control filters ensured reliable polarimetric measurements by requiring robust polarised detections ($\mathrm{SNR}_{\rm pol} > 4$), reliable RM values (\texttt{goodRM\_flag=True}), and well-fit Stokes-$I$ spectra, while blended sources, leakage-contaminated detections, and other instrumental artefacts were excluded.

Physically motivated Faraday rotation criteria were imposed to favour sources with intrinsic rather than Galactic foreground RM. Candidates were required to have moderate-to-high absolute rotation measures ($|{\rm RM}| > 400~\mathrm{rad~m^{-2}}$), significant RM excess relative to the local RM environment ($\Delta {\rm RM} = |{\rm RM} - \langle {\rm RM} \rangle_{\rm local}| > 50~\mathrm{rad~m^{-2}}$), and limited foreground turbulence ($\sigma_{\rm local~RM} < 500~\mathrm{rad~m^{-2}}$). Significant RM detections were further enforced by requiring a minimum RM significance of $2\sigma$.

Faraday-complex sources were prioritised using a ranking approach rather than strict cuts. The pipeline assigned additional weight to sources with high additional Faraday dispersion significance ($\sigma_{\rm add}/\delta \sigma_{\rm add}$), broadened Faraday spectra (${\rm RM}_{\rm width}/{\rm RMSF}_{\rm FWHM}$), and catalog complexity flags (\texttt{complex\_flag=True} and \texttt{complex\_M2\_CC\_flag=True}). Additional ranking preference was given to high polarised signal-to-noise, steep radio spectral indices, significant RM excess, and moderate source extension. This strategy effectively distinguished genuinely Faraday-complex systems from high-RM, Faraday-thin sources. For clarity, we have summarised the principal target selection criteria and thresholds discussed above in Table~\ref{tab:target_selection}.

Faraday complexity here refers to deviations from the simplest
Faraday-thin scenario, where all polarized emission originates from a
single component with one characteristic rotation measure (RM).
Complexity can arise when multiple polarized emitting regions with
different Faraday depths are present, when internal Faraday rotation
occurs within the emitting plasma, or when turbulent magneto-ionic
media introduce a distribution of RM values across the observed
emission. Such complexity produces broadened or asymmetric Faraday
dispersion spectra and wavelength-dependent depolarisation
\citep{Farnsworth_2011, Anderson_2015, OSullivan_2017}.

In this work, RM broadening refers to an observed Faraday spectrum that
is wider than the expected response of a Faraday-thin source. The
quantity $RM_{\rm width}/RM_{\rm RMSF,FWHM}$ compares the measured
Faraday-depth width of the source with the width of the rotation
measure spread function (RMSF), which represents the instrumental
Faraday resolution. Values significantly larger than unity indicate
that the source contains unresolved Faraday-depth structure beyond the
instrumental response.

\begin{table}[ht]
\centering
\caption{Target selection parameters for SPICE-RACS DR2 depolarisation study.}
\label{tab:target_selection}
\begin{tabular}{lll}
\hline
Parameter & Threshold & Description \\
\hline
$\mathrm{SNR}_{\rm pol}$ & $>4$ & Polarized S/N \\
goodRM\_flag & True & Reliable RM \\
Stokes-$I$ fit & Good & Reliable Stokes-I fit \\
$|{\rm RM}|$ & $>400$ & High absolute RM \\
$\Delta {\rm RM}$ & $>50$ & RM excess \\
$\sigma_{\rm local~RM}$ & $<500$ & Local RM scatter \\
RM significance & $>2\sigma$ & Significant RM \\
$\sigma_{\rm add}/\delta\sigma_{\rm add}$ & High & Faraday complexity \\
${\rm RM}_{\rm width}/{\rm RMSF}_{\rm FWHM}$ & High & Faraday-width excess \\ & & relative to RMSF\\
$\alpha$ & Steep & Spectral index \\
Angular size & Moderate & Source extension \\
\hline
\end{tabular}
\end{table}

Applying these criteria, we identified \texttt{RACS\_0900-28\_7036} as the highest-priority target. The source exhibits a large rotation measure,
$RM=345.7\pm0.2~{\rm rad\,m^{-2}}$, a substantial excess above the
local foreground ($\Delta RM\approx171~{\rm rad\,m^{-2}}$), significant
Faraday complexity ($\sigma_{\rm add}/\delta\sigma_{\rm add}\approx8.6$),
and a broadened Faraday spectrum
($RM_{\rm width}/RM_{\rm RMSF,FWHM}\approx1.26$), indicating that the
observed Faraday structure is wider than the instrumental response, confirming it is not a simple Faraday-thin emitter. 
The source additionally has a high polarised signal-to-noise ratio
($\mathrm{SNR}_{\rm pol}\approx172$), a steep spectral index
($\alpha\approx-0.71$), and an angular size of
$13.15''\times11.20''$. 
Since the SPICE-RACS DR2 angular resolution ranges from
$11.8''$ to $75.9''$,
the source is only marginally resolved at the highest-resolution end of the
survey and is effectively unresolved for a significant fraction of the ASKAP
observations. Therefore, the Faraday properties derived below should be
interpreted as effective properties of the polarized emission sampled within
the ASKAP beam rather than direct measurements of spatially resolved RM
variations. In particular, the inferred $\sigma_{\rm RM}$ values represent
beam-averaged and line-of-sight Faraday fluctuations, which may arise from
unresolved sub-beam structure, multiple polarized emitting regions, or
turbulent foreground magneto-ionic media. \citep{Anderson_2015, Anderson_2016, Pasetto_2018,
OSullivan_2017, Sun_2020, Sun_2025}.

\begin{figure}[htbp]
    \centering
    \includegraphics[width=0.49\textwidth]{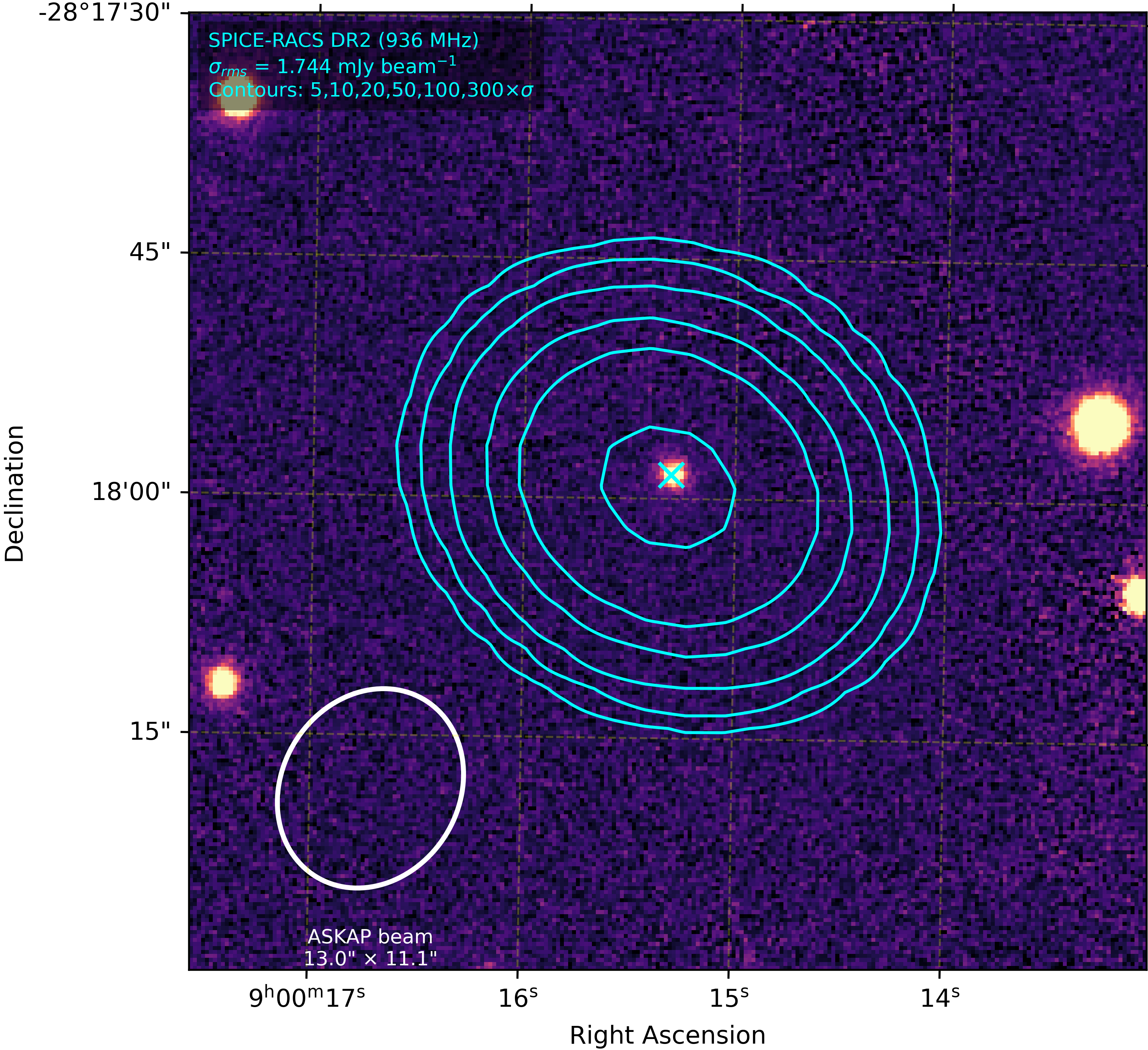}
    \caption{Optical--radio overlay of the ASKAP source 
    $\mathrm{cat\_id}=\texttt{RACS~0900-28\_7036}$. The background image shows 
    the optical emission from the Pan-STARRS1 $g$-band survey, while the radio 
    contours represent the SPICE-RACS DR2 Stokes-I emission from ASKAP. The 
    radio source is spatially coincident with the optical quasar 
    [VV98] J090015.4$-$281758 at a redshift of $z=0.9043\pm0.0061$. The source 
    was identified as a quasar based on optical spectroscopy (indicated by the cross mark), with the radio 
    counterpart confirmed from low-frequency and GHz radio surveys 
    \citep{Wenger2000,NED,Intema17,Condon98}.}
    \label{fig:RACS0900_radio_optical}
\end{figure}

In this study, we focus on the ASKAP source with 
\texttt{cat\_id}=RACS 0900-28\_7036. This is  
identified in the Pan-STARRS g-band image as a quasar [VV98] J090015.4$-$281758 (Figure~\ref{fig:RACS0900_radio_optical}). 
The source has a spectroscopic redshift of $z = 0.9043 \pm 0.0061$. The optical and 
radio positions are coincident within $<1^{\prime\prime}$, confirming the 
association between the radio emission and the quasar host 
\citep{Wenger2000, NED}. 

The source is also detected across multiple radio surveys, 
including the TGSS ADR at 150~MHz \citep{Intema17} and NVSS at 1.4~GHz \citep{Condon98}, where it has an 
integrated flux density of $S_{\rm 1.4\,GHz}=511.6 \pm 15.4$~mJy. A few key parameters of the selected source are listed in Table~\ref{tab:source_properties}.

\begin{table}[ht]
    \centering
    \caption{\textbf{SPICE-RACS DR2 Source: RACS\_0900$-$28\_7036}}
    \label{tab:source_properties}
    \begin{tabular}{l c}
    \hline
    \hline
    Right Ascension (J2000) & \(135.0638^\circ\) \\
    Declination (J2000) & \(-28.2998^\circ\) \\
    Galactic Latitude (\(b\)) & \(11.62^\circ\) \\
    Redshift ($z$) & 0.9 \\
    Rotation Measure (RM) & \(345.69 \pm 0.18\) rad m\(^{-2}\) \\
    RM Width\footnote{The RM width represents the FWHM of the Faraday dispersion function obtained
from RM synthesis. This quantity characterizes the width of the observed
Faraday spectrum and should not be confused with the QU-fitting parameter
$\sigma_{\rm RM}$, which describes the RM variance responsible for external
Faraday dispersion.} & \(77.68\) rad m\(^{-2}\) \\
    Polarized Intensity (P) & \(26.28 \pm 0.15\) mJy beam\(^{-1}\) \\
    Fractional Polarization (\(p\)) & \(3.83\%\) \\
    Stokes I Flux Density & \(686.96 \pm 0.23\) mJy beam\(^{-1}\) \\
    Total Integrated Flux Density & \(775.91 \pm 46.60\) mJy \\
    Spectral Index (\(\alpha\)) & \(-0.711 \pm 0.004\) \\
    Polarization Angle (\(\psi\)) & \(56.53^\circ \pm 0.17^\circ\) \\
    De-rotated Polarization Angle & \(6.68^\circ \pm 0.95^\circ\) \\
    Signal-to-noise Ratio (Polarized) & \(172.4\) \\
    Number of Channels & \(36\) \\
    Observed Frequency Range & \(803 - 1083\) MHz \\
    Nominal SPICE-RACS DR2 band & \(799.5 - 1087.5\) MHz \\
    Angular Size (Major Axis) & \(13.15''\) \\
    Angular Size (Minor Axis) & \(11.20''\) \\
    Faraday Complexity Flag & True \\
    Complexity Metric (\(\sigma_{\rm add}\)) & \(3.30 \pm 0.38\) \\
    Complex M2 CC Flag & True \\
    Morphology Code & M (multi-component) \\
    Telescope & ASKAP \\
    Survey & SPICE-RACS DR2 \\
    \hline
    \end{tabular}
\end{table}

\section{Faraday Rotation and Depolarisation}
\label{sec:polFar}
Polarized radio emission from active galaxies primarily originates from synchrotron radiation produced by relativistic electrons spiralling around magnetic field lines. As synchrotron emission is intrinsically linearly polarized, broadband polarimetric observations provide a powerful probe of the magneto-ionic environment within radio galaxies and along the line of sight \citep{Rybicki_Lightman_1979, Gaensler_2005}. The polarization properties of the radiation are described using the Stokes parameters $I$, $Q$, and $U$, where the complex linear polarization can be written as

\begin{equation}
P = Q + iU = p I e^{2 i \psi},
\end{equation}

where $p$ is the fractional polarization and $\psi$ is the polarization angle. The corresponding fractional Stokes parameters are defined as

\begin{equation}
q = \frac{Q}{I}, \qquad
u = \frac{U}{I},
\end{equation}

such that

\begin{equation}
p = \sqrt{q^2 + u^2},
\end{equation}

and the polarization angle is

\begin{equation}
\psi = \frac{1}{2}\tan^{-1}\left(\frac{u}{q}\right).
\end{equation}

As polarized radiation propagates through a magnetized ionized medium, the plane of polarization undergoes Faraday rotation \citep{Burn_1966, Sokoloff_1998, Rossetti_2008}. The magnitude of this rotation depends on both the thermal electron density and the line-of-sight magnetic field component. The corresponding Faraday depth is given by

\begin{equation}
\phi(s) = 0.81 \int_0^s n_e(s') B_\parallel(s') \, ds'
\quad \mathrm{rad\,m^{-2}},
\end{equation}

where \(n_e\) denotes the electron density (cm\(^{-3}\)), \(B_\parallel\) is the component of the magnetic field parallel to the line of sight (\(\mu\)G), and \(s\) represents the path length in parsecs. The integration is performed along the line of sight from the observer to the source \citep{Ferri_2021}. Under the simplest scenario of a single foreground Faraday screen, the observed polarization angle exhibits a linear dependence on the square of the wavelength, expressed as

\begin{equation}
\psi(\lambda^2) = \psi_0 + \mathrm{RM}\,\lambda^2,
\end{equation}

where $\psi_0$ denotes the intrinsic polarization angle, and RM represents the rotation measure.

In realistic astrophysical environments, the magneto-ionic medium is often turbulent and structurally complex, leading to wavelength-dependent depolarisation. This occurs when different regions within the source or telescope beam experience different amounts of Faraday rotation, causing partial cancellation of the polarized signal. Depolarisation can arise either internally, where the emitting and rotating media are mixed, or externally, where fluctuations originate in a foreground screen \citep{OSullivan_2012, OSullivan_2015, OSullivan_2017, Pasetto_2018, Passeto_2021, Taylor_2024}.

A simple Faraday-thin component with no depolarisation is described by

\begin{equation}
p(\lambda^2) =
p_0 \exp\left[2i\left(\psi_0 + \mathrm{RM}\lambda^2\right)\right],
\end{equation}

where $p_0$ is the intrinsic fractional polarization. In the presence of a turbulent foreground medium producing external Faraday dispersion, the polarized signal becomes

\begin{equation}
p(\lambda^2) =
p_0
\exp\left[2i\left(\psi_0 + \mathrm{RM}\lambda^2\right)\right]
\exp\left(-2\sigma_{\mathrm{RM}}^2 \lambda^4\right),
\end{equation}

where $\sigma_{\mathrm{RM}}$ characterizes the RM fluctuations across the source or beam. Internal depolarisation, often referred to as a Burn slab or Faraday-thick component \citep{Burn_1966, Sokoloff_1998, Pasetto_2018}, can be expressed as

\begin{equation}
p(\lambda^2) =
p_0
\exp\left[2i\left(\psi_0 + \frac{1}{2}R\lambda^2\right)\right]
\frac{\sin(R\lambda^2)}{R\lambda^2},
\end{equation}

where $R$ is the Faraday depth extent through the emitting region.

The wide fractional bandwidth and dense spectral sampling of SPICE-RACS DR2 make it particularly sensitive to Faraday-complex behaviour and wavelength-dependent depolarisation. The survey covers $\lambda^2 \approx 0.076$--$0.14~\mathrm{m}^2$, corresponding to observing frequencies of $799.5$--$1087.5$ MHz. For an external Faraday dispersion model, the fractional polarization scales as

\begin{equation}
|p(\lambda^2)| =
p_0 \exp\left(-2\sigma_{\mathrm{RM}}^2 \lambda^4\right).
\end{equation}

Across the ASKAP/SPICE-RACS band, this implies that sources with $\sigma_{\mathrm{RM}} \gtrsim 5$--$10~\mathrm{rad\,m^{-2}}$ already exhibit substantial depolarisation toward the lower-frequency end of the band, while systems with $\sigma_{\mathrm{RM}} \gtrsim 15$--$20~\mathrm{rad\,m^{-2}}$ can become strongly depolarized across much of the observing window. Consequently, SPICE-RACS is highly effective at identifying magnetized radio galaxies with complex Faraday structures and significant internal or external depolarising media \citep{Anderson_2021, Anderson_2023, Gaensler_2025, Gaensler_2025_CGM}.

\subsection{Depolarization Models}
\label{sec:depolmodels}
Broadband ASKAP spectropolarimetric observations are sensitive to both astrophysical Faraday complexity and low-level instrumental polarization leakage. Residual instrumental polarization is typically concentrated near ${\rm RM}\approx0~{\rm rad~m^{-2}}$, but can broaden due to ionospheric RM fluctuations and calibration imperfections \citep{Jelic_2014, Mevius_2018, Van_Eck_2025}. To avoid contamination of the intrinsic source polarization, the instrumental contribution must therefore be included explicitly in the depolarization modelling.

In order to model the observed Stokes $Q$ and $U$ spectra of the source, we considered combinations of Faraday-thin, Burn-slab (Faraday-thick), and external Faraday dispersion components \citep{Pirasthesis_2024}. The complex polarization for a Faraday-thin component is given by

\begin{equation}
p_T(\lambda^2;p_0,\psi_0,RM)
=
p_0\,e^{2i(\psi_0+RM\lambda^2)},
\label{eq:thin}
\end{equation}

where $p_0$ is the intrinsic fractional polarization, $\psi_0$ is the intrinsic polarization angle, and $RM$ is the rotation measure.

External Faraday dispersion (EFD), representing depolarization produced by a turbulent foreground magneto-ionic medium \citep{Goodlet_2004, Laing_2008, OSullivan_2017, OSullivan_2018, Brentjens_2019}, is modelled as

\begin{equation}
p_{\rm EFD}(\lambda^2;p_0,\psi_0,RM,\sigma_{\rm RM})
=
p_0\,e^{2i(\psi_0+RM\lambda^2)}
e^{-2\sigma_{\rm RM}^2\lambda^4},
\label{eq:efd}
\end{equation}

where $\sigma_{\rm RM}$ describes the Faraday dispersion within the external screen.

It is important to note that $\sigma_{\rm RM}$ in the external
Faraday dispersion model does not necessarily require the source to be
spatially resolved. Instead, it represents the variance of Faraday
depths sampled by the observed polarized emission within the telescope
beam and along the line of sight. For a purely point-like
Faraday-thin source observed through a uniform foreground screen,
$\sigma_{\rm RM}$ would approach zero, and the depolarization term $\exp(-2\sigma_{\rm RM}^{2}\lambda^{4})$
would become unity. Equations~8, 10, and 12 would then reduce to the
standard Faraday-thin rotation law. However, unresolved internal
structure, multiple polarized components, or turbulent foreground
screens on scales smaller than the restoring beam can still produce a
non-zero effective $\sigma_{\rm RM}$. Therefore, the measured $\sigma_{\rm RM}$ values in this work should be
interpreted as effective RM dispersions produced by unresolved Faraday
structure and/or line-of-sight fluctuations, rather than evidence for
resolved RM variations across the radio source.

A Burn-slab (Faraday-thick) component is described by

\begin{equation}
p_{\rm slab}(\lambda^2;p_0,\psi_0,R)
=
p_0\,e^{2i(\psi_0+\frac{1}{2}R\lambda^2)}
\frac{\sin(R\lambda^2)}{R\lambda^2},
\label{eq:slab}
\end{equation}

where $R$ characterizes the Faraday depth extent of the slab \citep{Burn_1966, Sokoloff_1998}.

Using these functional forms, we explored five depolarization configurations:

\begin{itemize}
    \item \textbf{m1:} a single external Faraday dispersion component (1 EFD),
    \item \textbf{m2:} one Faraday-thin instrumental component plus one external Faraday dispersion component (1 Thin + 1 EFD),
    \item \textbf{m3:} one Faraday-thin instrumental component plus two external Faraday dispersion components (1 Thin + 2 EFD),
    \item \textbf{m4:} one Burn-slab instrumental component plus one external Faraday dispersion component (1 Slab + 1 EFD),
    \item \textbf{m5:} one Burn-slab instrumental component plus two external Faraday dispersion components (1 Slab + 2 EFD).
\end{itemize}

For the ``Thin'' models (m2 and m3), the residual instrumental polarization leakage was represented using a Faraday-thin component constrained near ${\rm RM}\approx0~{\rm rad~m^{-2}}$. For the ``Slab'' models (m4 and m5), the instrumental contribution was instead modelled using a Burn-slab component, allowing for broadened leakage structures arising from ionospheric RM fluctuations across the ASKAP observing band. The astrophysical polarized emission was modelled using one or two external Faraday dispersion components to account for turbulent magneto-ionic media and multiple Faraday-active regions along the line of sight.

These five models were compared using Bayesian evidence and goodness-of-fit statistics in order to determine the preferred physical description of the polarized source.

\section{QU-fitting and Bayesian model selection}
\label{sec:methods}
For the ASKAP QU-fitting analysis, physically motivated priors were applied to guide the parameter exploration while ensuring realistic polarization and Faraday rotation behaviour. Instrumental polarization was modelled either as a Faraday-thin or Burn-slab component, with the Rotation Measure for the thin component constrained to $-3 \leq {\rm RM}_{\rm leak,thin} \leq 1.5~{\rm rad~m^{-2}}$ and for the thick component to $-6 \leq {\rm RM}_{\rm leak,thick} \leq 3~{\rm rad~m^{-2}}$. The fractional polarization of all components was restricted to physically plausible values $0 \leq p_0 \leq 1$, and the polarization angles were limited to $0^\circ \leq \psi_0 \leq 180^\circ$. For external Faraday-dispersing components, the Faraday dispersion $\sigma_{\rm RM}$ was assigned a uniform prior from $0$ to $30~{\rm rad~m^{-2}}$, reflecting the sensitivity of ASKAP to broad RM structures while remaining agnostic about the detailed substructure \citep{LOFAR_2023}. This range for $\sigma_{\rm RM}$ reflects ASKAP's capability to probe both moderately turbulent and strongly Faraday-broadened emission, which would be heavily depolarized at lower frequencies such as those probed by LOFAR \citep{LOFAR_2023}. The Rotation Measure was assigned a broad uniform prior within $-1000 \leq {\rm RM} \leq 1000~{\rm rad~m^{-2}}$ to capture potential complex Faraday structures. These priors ensure robust separation of intrinsic source polarization from instrumental leakage and allow the model-fitting algorithm to explore the relevant Faraday complexity without imposing overly restrictive assumptions.

To investigate the Faraday structure and depolarization behaviour of the selected source, we applied one-dimensional QU-fitting to the SPICE-RACS DR2 spectropolarimetric data. The fitting was performed on the fractional Stokes parameters,

\begin{equation}
q(\lambda^2)=\frac{Q}{I}, \qquad
u(\lambda^2)=\frac{U}{I},
\end{equation}

which reduces sensitivity to uncertainties in the absolute flux-density scale and Stokes-$I$ modelling. Consequently, model selection is primarily driven by the spectral behaviour of $q(\lambda^2)$ and $u(\lambda^2)$, making the analysis robust to calibration-related amplitude uncertainties while remaining highly sensitive to Faraday rotation and depolarization signatures \citep{OSullivan_2012, OSullivan_2017, Paul_2026}.

The qu-fitting analysis was performed using the \texttt{RM-Tools} software package\footnote{\url{https://github.com/CIRADA-Tools/RM-Tools}} \citep{Purcell_2020, Van_Eck_2026}, which employs the \texttt{PyMultiNest} interface to the \texttt{MultiNest} nested-sampling algorithm \citep{Feroz_2008, Feroz_2009, Feroz_2019}. For each depolarization model, the algorithm explores the multidimensional parameter space, determines the posterior probability distributions of the model parameters, and computes the Bayesian evidence ($Z$), which quantifies the probability of the observed data given the model \citep{Pirasthesis_2024, Paul_2026}.

Model comparison was performed using the logarithmic Bayes factor,

\begin{equation}
\Delta \ln Z = \ln Z_a - \ln Z_b
             = \ln \left(\frac{Z_a}{Z_b}\right),
\end{equation}

where \(Z_a\) and \(Z_b\) are the Bayesian evidences of two competing models. The relative support for each model was quantified using \(2\Delta \ln Z\), with the classification scheme of \citet{Kass_1995}. Specifically, \(2\Delta \ln Z < 2\) is regarded as inconclusive, \(2 \leq 2\Delta \ln Z < 6\) as positive evidence, \(6 \leq 2\Delta \ln Z < 10\) as strong evidence, and \(2\Delta \ln Z \geq 10\) as very strong evidence in favour of the model with the higher Bayesian evidence.

In addition to the Bayesian evidence, the goodness-of-fit of each model was evaluated using the reduced chi-squared statistic,

\begin{equation}
\chi^2_{\rm red}
=
\frac{1}{\rm DoF}
\sum_{i=1}^{N}
\left[
\left(
\frac{q_i-q_{{\rm model},i}}
{\sigma_{q_i}}
\right)^2
+
\left(
\frac{u_i-u_{{\rm model},i}}
{\sigma_{u_i}}
\right)^2
\right],
\end{equation}

where DoF ($N - N_{\rm free}$) is the number of degrees of freedom for the model, $N$ is the number of frequency channels used in the fit, $q_{\rm model}$ and $u_{\rm model}$ are the model predictions, and $\sigma_q$ and $\sigma_u$ are the corresponding uncertainties. The reduced chi-squared provides an independent assessment of fit quality while accounting for the number of free parameters in the model.

The five depolarization models described in Section~\ref{sec:depolmodels} were fitted to the ASKAP polarization spectra. Bayesian model comparison identified the \texttt{m5} model, consisting of an instrumental Burn-slab component and two external Faraday dispersion components, as the preferred representation of the data with the highest Bayesian evidence ($\ln Z = 280.23$). The alternative \texttt{m3} model, which models the instrumental polarization as a Faraday-thin component together with two external Faraday dispersion components, yielded a comparable evidence ($2\Delta\ln Z = 2.4$), indicating that the current data do not strongly distinguish between the two descriptions of the instrumental contribution \citep{Pirasthesis_2024}. In contrast, the simpler one- and two-component models (\texttt{m1}, \texttt{m2}, and \texttt{m4}) are strongly disfavoured ($2\Delta\ln Z > 25$). The preference for the multi-component models demonstrates that the polarized emission from \texttt{RACS\_0900$-$28\_7036} cannot be adequately described by a single Faraday-rotating screen and instead requires multiple Faraday-active components with significant Faraday dispersion.

\begin{table*}[ht]
\centering
\caption{Summary of the Bayesian model comparison for the ASKAP SPICE-RACS source 
RACS\_0900$-$28\_7036.}
\label{tab:modelcomparison}
\begin{tabular}{lcccccc}
\hline
Model & $N_{\rm free}$ & DoF & $\chi^2_{\rm red}$ & $\ln Z$ & $2\Delta \ln Z$ & Interpretation \\
\hline
m5 (1 Slab + 2 EFD) & 11 & 60 & 2.93 & 280.23 & 0.00 & Preferred \\
m3 (1 Thin + 2 EFD) & 11 & 60 & 4.11 & 279.03 & 2.40 & Comparable \\
m2 (1 Thin + 1 EFD) & 7 & 64 & 2.55 & 267.31 & 25.84 & Strongly disfavoured \\
m4 (1 Slab + 1 EFD) & 7 & 64 & 2.55 & 266.64 & 27.18 & Strongly disfavoured \\
m1 (1 EFD) & 4 & 67 & 4.52 & 207.00 & 146.46 & Strongly disfavoured \\
\hline
\end{tabular}
\end{table*}

\section{Results}
\label{sec:results}
We analyse the depolarization behaviour of the
ASKAP source RACS 0900-28 7036 using multi-component
Faraday models. The best-fitting model,
based on both the Bayesian evidence and reduced chi-squared
($\chi^2_{\rm red}$), is \texttt{m5}, which includes one Burn-slab instrumental component and two external Faraday
dispersion (EFD) components (1 Slab + 2 EFD).
This model accounts for a minor instrumental Burn-slab contribution at
$R_{\rm slab} \approx -3.25~{\rm rad~m^{-2}}$
and two external Faraday dispersion (EFD) components centred at
$RM_{\rm EFD,1} \approx 345.5~{\rm rad~m^{-2}}$ and $RM_{\rm EFD,2} \approx 131.5~{\rm rad~m^{-2}}$ representing distinct magneto-ionic structures along the line of sight. 
 A summary of the model comparison is presented in Table~\ref{tab:modelcomparison}.

\begin{figure*}[!htbp]
    \centering
    \includegraphics[width=0.99\textwidth]{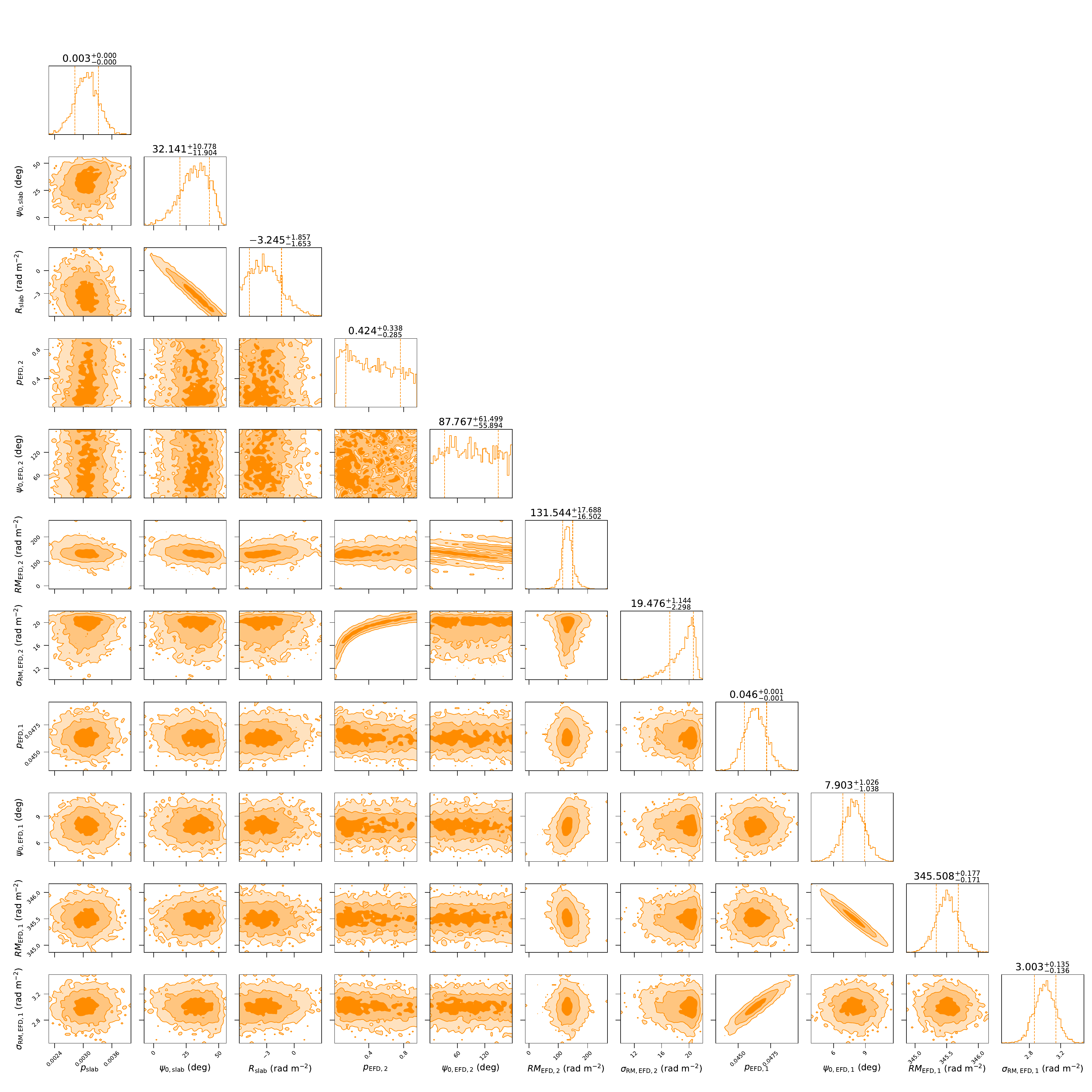}
    \caption{Posterior distributions and parameter covariances for the preferred 
    \texttt{m5} polarization model obtained from the broadband $qu$-fitting. The diagonal panels present the marginalized distributions of each model parameter, together with their corresponding $1\sigma$ (68\%) confidence intervals. The off-diagonal panels display the joint posterior distributions for pairs of parameters, illustrating correlations and degeneracies within the model parameter space. Contours enclosing 68\%, 95\%, and 99.7\% of the posterior probability are overlaid to indicate the $1\sigma$, $2\sigma$, and $3\sigma$ confidence regions, respectively.}
    \label{fig:corner}
\end{figure*}

The posterior distributions shown in Figure~\ref{fig:corner} indicate that several model parameters are correlated, as expected for broadband qu-fitting of Faraday-complex sources. In particular, the intrinsic fractional polarization $p_0$, intrinsic polarization angle $\chi_0$, and Faraday dispersion parameters exhibit partial degeneracies, since different combinations of these quantities can reproduce similar Stokes $q$ and $u$ spectra over a finite frequency band. The dominant RM component, however, is comparatively well constrained, with a narrow posterior centred near the catalogued value, indicating that the main Faraday rotation is robustly recovered by the data. By contrast, the weaker and broader component is less tightly constrained, and its exact decomposition may depend modestly on the assumed priors and model parametrization.

\subsection{Best-fit polarization components}
The first external Faraday dispersion (EFD) component is centred at
\[
{\rm RM} = 345.50 \pm 0.17 \ {\rm rad \ m^{-2}},
\]
consistent with the catalogued SPICE-RACS RM value of
\[
345.69 \pm 0.18 \ {\rm rad \ m^{-2}}.
\]
This component exhibits a relatively modest Faraday dispersion of
\[
\sigma_{\rm RM} \approx 3.0 \ {\rm rad \ m^{-2}},
\]
indicating only mild fluctuations in the foreground magneto-ionic medium. We compared the measured RM with the Galactic foreground rotation measure at the source position\footnote{\url{https://wwwmpa.mpa-garching.mpg.de/~ensslin/research/data/faraday2020.html}} \citep{Hutschenreuter22}, finding
\[
{\rm RM}_{\rm Gal} = 331.89 \pm 33.05 \ {\rm rad \ m^{-2}},
\]
which is fully consistent with the dominant RM component within the uncertainties. This suggests that the principal Faraday rotation is largely accounted for by the Galactic foreground, with only a relatively small residual contribution from the immediate environment of the radio source or other intervening magnetized structures.

A second EFD component is detected at
\[
{\rm RM} \approx 132 \ {\rm rad \ m^{-2}}
\]
with a substantially larger Faraday dispersion of
\[
\sigma_{\rm RM} \approx 19.5 \ {\rm rad \ m^{-2}}.
\]
This component is comparatively poorly constrained, particularly in intrinsic polarization angle and fractional polarization, indicating significant degeneracy in the fit. Physically, such a large dispersion implies a highly turbulent or Faraday-thick magnetized medium capable of strongly depolarizing the polarized emission across the ASKAP observing band. This
reflect the high dynamic range and broad Faraday sensitivity of ASKAP, allowing the detection of both moderately and strongly depolarized components along the line of sight \citep{Gaensler_2004, Anderson_2021, Anderson_2024, Gaensler_2025}.

The Burn-slab component has a small intrinsic polarized fraction,
\[
p_0 \approx 0.003,
\]
and a Faraday depth extent of
\[
R_{\rm slab} \approx -3.3 \ {\rm rad \ m^{-2}},
\]
which is close to zero Faraday depth. This weak component likely represents residual low-level instrumental leakage. The fitted fractional polarizations are
$p_{\rm slab}=0.00307$,
$p_{\rm EFD,1}=0.0463$, and 
$p_{\rm EFD,2}=0.4239$,
indicating that the bulk of the polarized emission is captured by the
first EFD component.
A summary of the best fit model parameters are given in Table~\ref{tab:m5params}. 
The dominant EFD component associated with 
${\rm RM}\approx345~{\rm rad~m^{-2}}$ is tightly constrained, with narrow posterior 
distributions in both RM and $\sigma_{\rm RM}$. 
This indicates that the broadband ASKAP coverage provides strong leverage for 
recovering the principal Faraday rotating component \citep{Thomson_2026}.

\begin{table}[!htbp]
\centering
\caption{Best-fitting parameters for the preferred \texttt{m5} polarization model. The model consists of one instrumental Burn-slab component and two external Faraday dispersion (EFD) components.}
\label{tab:m5params}
\begin{tabular}{lccc}
\hline
Parameter & Median & $+1\sigma$ & $-1\sigma$ \\
\hline
$p_{\rm slab}$ & 0.0031 & 0.0002 & 0.0002 \\
$\psi_{0,\rm slab}$ (deg) & 32.1 & 10.8 & 11.9 \\
$R_{\rm slab}$ (rad m$^{-2}$) & $-3.25$ & 1.86 & 1.65 \\
\hline
$p_{\rm EFD,1}$ & 0.0463 & 0.0009 & 0.0008 \\
$\psi_{0,\rm EFD,1}$ (deg) & 7.90 & 1.03 & 1.04 \\
${\rm RM}_{\rm EFD,1}$ (rad m$^{-2}$) & 345.5 & 0.18 & 0.17 \\
$\sigma_{\rm RM,EFD,1}$ (rad m$^{-2}$) & 3.00 & 0.13 & 0.14 \\
\hline
$p_{\rm EFD,2}$ & 0.424 & 0.338 & 0.285 \\
$\psi_{0,\rm EFD,2}$ (deg) & 87.8 & 61.5 & 55.9 \\
${\rm RM}_{\rm EFD,2}$ (rad m$^{-2}$) & 131.5 & 17.7 & 16.5 \\
$\sigma_{\rm RM,EFD,2}$ (rad m$^{-2}$) & 19.5 & 1.14 & 2.30 \\
\hline
\end{tabular}
\end{table}

The environment of RACS~0900$-$28~7036 was further investigated
using X-ray data. The source is detected in the
eROSITA-DE DR1 $0.2$--$2.3$~keV band, with the X-ray emission spatially
coincident with the optical/radio counterpart. The compact X-ray
morphology is consistent with emission associated with the active
galactic nucleus. Inspection of the wider $12'\times12'$ X-ray field
does not reveal significant diffuse emission extending on arcminute
scales that would be expected from a massive galaxy cluster or a
bright intracluster medium \citep{Merloni2024}. Therefore, although a
lower-mass group environment cannot be ruled out, the available
X-ray data do not support the presence of a rich cluster environment.
The external RM components identified from the depolarisation analysis
are therefore more likely associated with smaller-scale magnetised
structures, such as the quasar host environment, group-scale halos,
intervening galaxies, or other ionised structures along the line of
sight \citep{AkahoriRyu2011, Bonafede2010, XuHan2014}.

\subsection{Polarization spectra}
\begin{figure*}
    \centering
    \includegraphics[width=\textwidth]{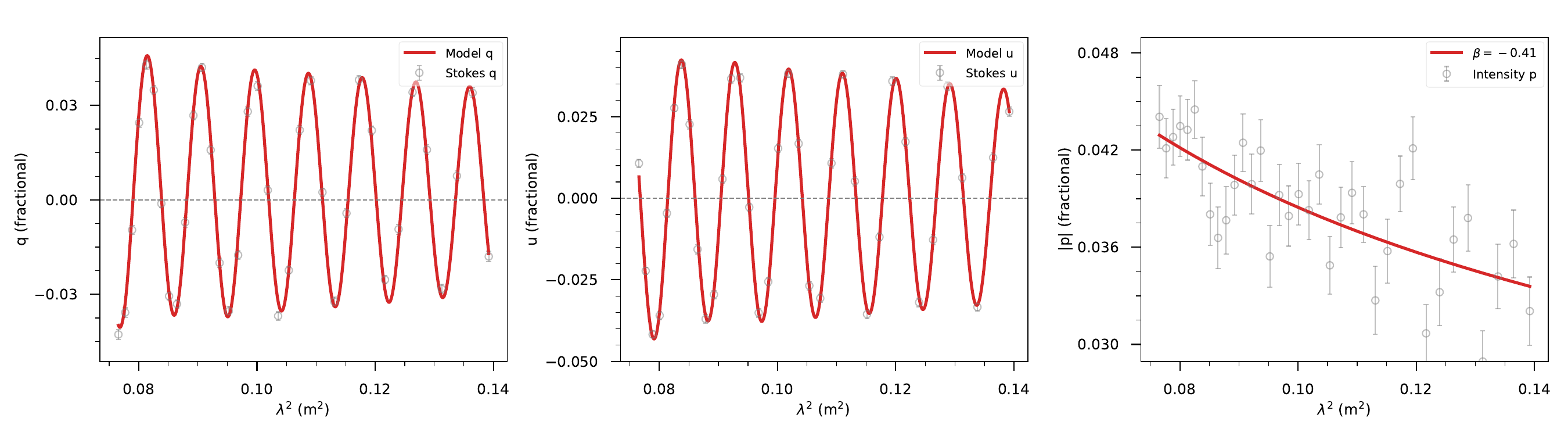}
    \caption{Fractional Stokes $q$, $u$, and polarized fraction $|p|$ as a function 
    of $\lambda^2$, together with the best-fitting \texttt{m5} model. 
    The decreasing oscillation amplitude towards larger $\lambda^2$ indicates significant Faraday depolarization. The solid red line in the right panel shows the best-fitting power-law model, \(|p|=A(\lambda^2)^\beta\), yielding \(\beta=-0.41\) and implying moderate depolarization across the ASKAP band.}
    \label{fig:pqufit}
\end{figure*}

The observed fractional Stokes $q$, $u$, and polarized fraction $|p|$ spectra, 
together with the best-fitting model, are shown in 
Figure~\ref{fig:pqufit}. 
The oscillatory behaviour observed in both $q$ and $u$ as a function of 
$\lambda^2$ is a direct signature of Faraday rotation. 
However, the amplitude of the oscillations gradually decreases towards longer 
wavelengths, indicating significant depolarization across the ASKAP band.

The decline in polarized fraction with increasing $\lambda^2$ is consistent with 
external Faraday dispersion produced by turbulent magnetized plasma. 
In the Burn-law formalism \citep{Burn_1966}, the polarized fraction decreases as
\[
|p(\lambda^2)| = p_0 \exp(-2\sigma_{\rm RM}^2\lambda^4),
\]
so even moderate values of $\sigma_{\rm RM}$ can strongly suppress polarization 
at long wavelengths. 
For the dominant component, the inferred 
$\sigma_{\rm RM}\sim3~{\rm rad~m^{-2}}$ is sufficient to produce measurable 
depolarization within the ASKAP frequency range. 
The second broad component with 
$\sigma_{\rm RM}\sim20~{\rm rad~m^{-2}}$ would depolarize much more rapidly, 
consistent with its weak observational constraints.

We also show the fractional polarized intensity, \(|p|=\sqrt{q^2+u^2}\), as a function of \(\lambda^2\), together with the corresponding best-fitting power-law model, $|p| = A\,(\lambda^2)^{\beta}$. The best-fitting slope, ($\beta=-0.41\pm0.09$), suggests the presence of moderate depolarization across the ASKAP band. Although the polarized intensity spectrum shows a smooth monotonic decline as a function of wavelength, the associated q and u spectra exhibit substantially greater Faraday complexity. This demonstrates that polarized intensity alone may not fully capture the underlying magneto-ionic structure, emphasizing the need for comprehensive spectro-polarimetric modelling of ASKAP polarization data \citep{OSullivan_2017, Taylor_2024, Sinha_2026}.

\subsection{Faraday Rotation of the polarization angle}

\begin{figure}[!htbp]
    \centering
    \includegraphics[width=0.48\textwidth]{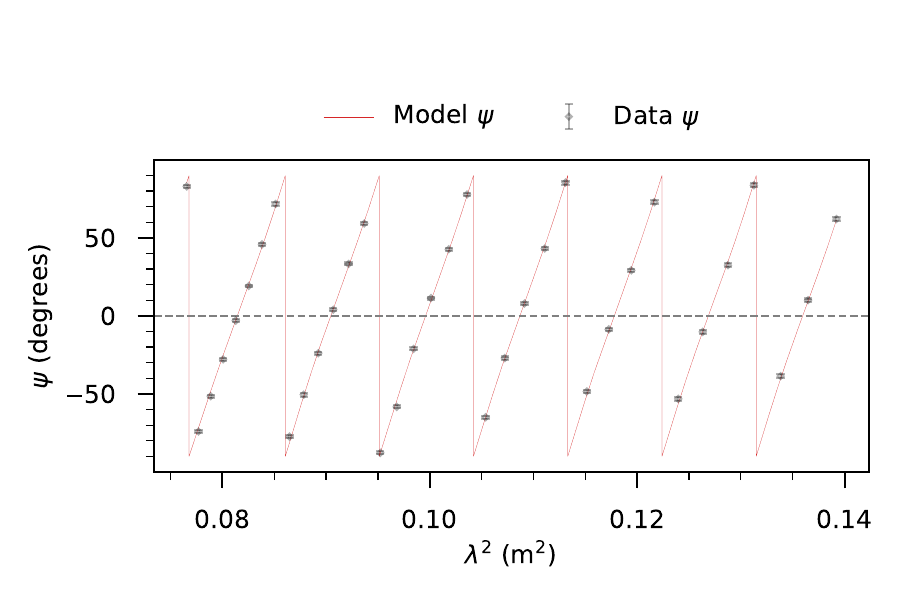}
    \caption{Polarization angle as a function of $\lambda^2$ for RACS\_0900-28\_7036 with the best-fitting m5 model.}
    \label{fig:psifit}
\end{figure}

The polarization angle $\psi$ as a function of $\lambda^2$ is shown in Figure~\ref{fig:psifit}. Across the ASKAP band, $\psi$ exhibits an approximately linear dependence on $\lambda^2$, indicating that the dominant rotation is well described by a foreground Faraday screen. At the same time, the residual deviations from linearity imply that the full polarized emission may include additional complexity beyond a single Faraday-thin component. Consequently, the polarization-angle behaviour should be interpreted as supportive, but not conclusive, evidence for a simple Faraday structure.

\begin{figure}[!htbp]
    \centering
    \includegraphics[width=0.45\textwidth]{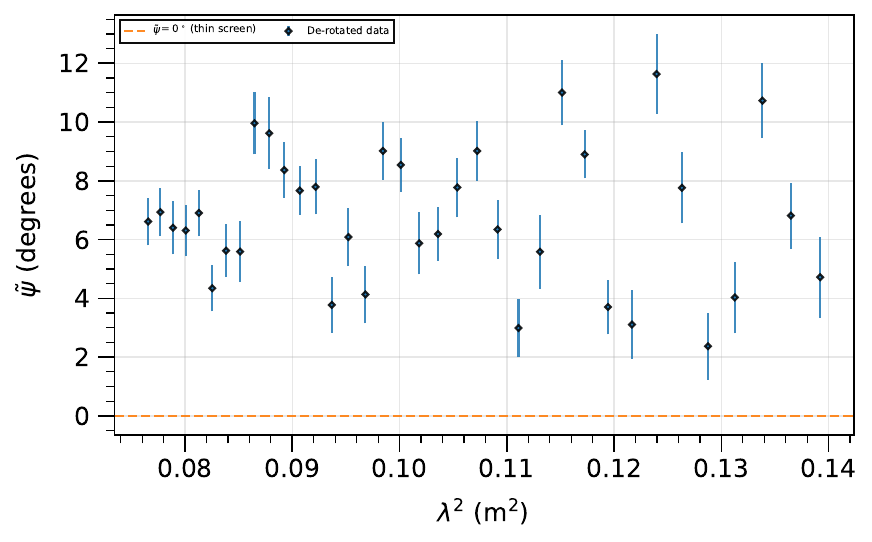}
    \caption{De-rotated polarization angle spectrum for RACS\_0900-28\_7036 after removing the catalogue RM contribution. For a single Faraday-thin screen, the de-rotated angles would be expected to scatter symmetrically around the orange dashed line ($\tilde{\psi}=0^\circ$). }
    \label{fig:psiderot}
\end{figure}

The de-rotated polarization angle spectrum, shown in Figure~\ref{fig:psiderot}, was obtained by unwrapping the observed angles and removing the catalogue RM component:
\[
\tilde{\psi}(\lambda^2) = \psi(\lambda^2) - {\rm RM}_{\rm cat} \lambda^2,
\]
followed by $\pm180^\circ$ unwrapping. For a single Faraday-thin screen at ${\rm RM}_{\rm cat} \approx 345.7~{\rm rad~m^{-2}}$, the de-rotated angles would be expected to scatter symmetrically around $\tilde{\psi}=0^\circ$. Instead, the de-rotated polarization angles remain systematically offset above
zero, clustering around $\tilde{\psi}\approx7^\circ$. This value is consistent
with the intrinsic polarization angle of the dominant Faraday-dispersive
component in the preferred m5 model
($\psi_{0,\mathrm{EFD,1}} = 7.90^{+1.03}_{-1.04}\,^\circ$), indicating that
the catalogue RM successfully removes the dominant Faraday rotation while
leaving the intrinsic polarization angle of the source unchanged.

A linear fit to the de-rotated angles yields a residual rotation measure ${\rm RM}_{\rm res}=-0.06 \pm 20.87~{\rm rad~m^{-2}}$, statistically consistent with zero, and a quadratic term $(-7.33 \pm 12.5 \times 10^2)~{\rm deg~m^{-4}}$ that is not strongly significant. The reduced chi-squared value is $\chi^2_\mathrm{red} = 4.67$, indicating that a simple linear $\tilde{\psi} \propto \lambda^2$ relation captures the majority of the polarization angle behaviour.

The slight positive offset in $\tilde{\psi}$ reflects the intrinsic polarization angle of the dominant Faraday component. While the angle spectrum appears largely Faraday-simple, the full $q$/$u$ depolarization modeling reveals additional Faraday complexity associated with weaker components at RM $\approx 132~{\rm rad~m^{-2}}$ and the instrumental Burn-slab near RM $\approx 0~{\rm rad~m^{-2}}$, which are captured by the preferred \texttt{m5} model.

\subsection{Polarization behaviour in the q-u plane}

\begin{figure}[!htbp]
    \centering
    \includegraphics[width=0.45\textwidth]{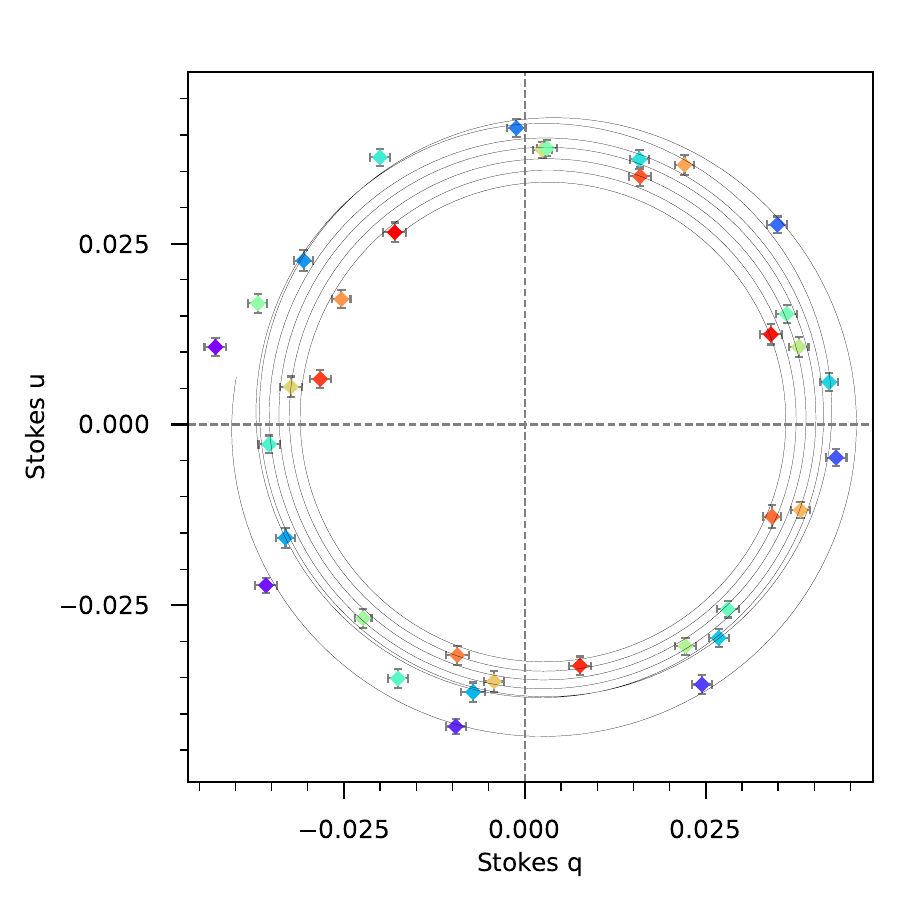}
    \caption{Evolution of the polarization vector in the complex q-u plane for the ASKAP source RACS\_0900-28\_7036, coloured by observing frequency (blue = high frequency / short $\lambda^2$, red = low frequency / long $\lambda^2$). 
Data points include $1\sigma$ error bars, and the solid line shows the best-fit m5 model (1 Slab + 2 EFD). 
The gradual contraction of the polarization vector amplitude from blue to red illustrates the wavelength-dependent depolarization captured by the model.}
    \label{fig:quplane}
\end{figure}

The trajectory of the polarization vector in the complex $q-u$ plane is shown in 
Figure~\ref{fig:quplane}. 
For a purely Faraday-thin source, the polarization vector would trace a circular 
path with constant amplitude. 
Instead, the source exhibits a spiral-like trajectory with decreasing amplitude, 
demonstrating that depolarization progressively suppresses the polarized signal at 
longer wavelengths.

Such behaviour is characteristic of Faraday-complex sources containing multiple 
polarized components or turbulent foreground screens. 
The shrinking radius of the trajectory in the $q-u$ plane reflects the
loss of coherent polarization due to Faraday dispersion. Since the
source is only marginally resolved at SPICE-RACS DR2 resolution, this
effect should be interpreted as arising from unresolved Faraday
structure and/or beam-averaged RM fluctuations rather than necessarily
from resolved spatial variations across the radio source \citep{Burn_1966, Brentjens_2005, Sotomayor_2008}.

Overall, the ASKAP broadband polarization data reveal that RACS\_0900$-$28\_7036 is a Faraday-complex radio source whose polarized emission arises from multiple magneto-ionic components along the line of sight. Although the polarization-angle spectrum is largely consistent with a dominant Faraday-rotating component, the broadband $q$ and $u$ behaviour requires additional depolarizing structures to reproduce the observed polarization properties. This demonstrates that a single rotation measure is insufficient to fully describe the source. The presence of multiple Faraday-dispersive components suggests a structured and inhomogeneous magnetized environment, likely containing regions with different magnetic-field strengths, electron densities, and levels of turbulence \citep{OSullivan_2017, Pasetto_2018, Passeto_2021}. These results highlight the power of broadband ASKAP spectropolarimetry for resolving complex Faraday structures and probing the magneto-ionic conditions surrounding radio galaxies.

\section{Summary and Conclusions}
\label{sec:discussion}
We have presented a broadband spectropolarimetric analysis of the SPICE-RACS DR2 source \texttt{RACS\_0900$-$28\_7036}, selected from a newly developed target-selection pipeline designed to identify highly polarized and Faraday-complex radio sources in the Rapid ASKAP Continuum Survey (RACS-low3) polarization catalogue. The source was chosen on the basis of its large rotation measure, significant excess above the local foreground RM, high polarized signal-to-noise ratio, and multiple catalog indicators of Faraday complexity.

Using broadband $qu$-fitting and Bayesian model comparison, we explored a suite of depolarization models incorporating instrumental polarization leakage and astrophysical Faraday-dispersive components. The preferred model (\texttt{m5}) consists of a weak instrumental Burn-slab component together with two external Faraday dispersion (EFD) components. Although the alternative \texttt{m3} model provides a statistically comparable description of the data, both favoured models require multiple Faraday-active components, demonstrating that the polarized emission cannot be explained by a single Faraday-thin screen. The detected Faraday complexity does not necessarily imply that the
source is spatially resolved. Instead, it indicates that the polarized
emission contains multiple Faraday-depth components or RM variations
within the effective ASKAP beam and along the line of sight.

The dominant polarized component is centred at ${\rm RM}\approx345.5~{\rm rad~m^{-2}}$, in excellent agreement with the catalogue RM, and exhibits modest Faraday dispersion ($\sigma_{\rm RM}\approx3~{\rm rad~m^{-2}}$). 
The second EFD component at
$RM\approx132~{\rm rad\,m^{-2}}$
displays substantially larger Faraday dispersion
($\sigma_{\rm RM}\approx19.5~{\rm rad\,m^{-2}}$),
indicating the presence of a more turbulent or Faraday-thick
magneto-ionic environment. Although, the source is not spatially resolved by most SPICE-RACS DR2
observations, this non-zero dispersion does not imply resolved RM structure.
Instead, it traces unresolved Faraday complexity within the effective ASKAP
beam and along the propagation path.
The polarization-angle spectrum is largely consistent with the dominant RM component, while the broadband $q$ and $u$ behaviour reveals additional complexity that is not evident from a simple rotation-measure analysis alone. The polarization vector trajectory in the $q-u$ plane exhibits the expected spiral contraction, and the fitted polarization spectral index $\beta=-0.41\pm0.09$ quantifies the observed depolarization across the ASKAP band.

The source therefore represents a clear example of a Faraday-complex radio galaxy whose polarized emission probes multiple magnetized structures along the line of sight. The combination of a large RM, significant depolarization, and multiple Faraday-dispersive components suggests an inhomogeneous magneto-ionic environment containing regions with differing magnetic-field strengths, electron densities, and turbulence levels. These results demonstrate the diagnostic power of broadband ASKAP spectropolarimetry for disentangling complex Faraday structures and characterizing the magnetized environments of radio galaxies.

Looking ahead, the methodology developed in this work will be applied to the full SPICE-RACS DR2 polarization catalogue. Systematic broadband depolarization modelling of many thousands of ASKAP polarized sources will enable statistical studies of Faraday complexity across different source populations and environments. Such analyses will provide new constraints on the origin of extreme rotation measures, the prevalence of turbulent magnetized plasma in radio-galaxy environments, and the evolution of cosmic magnetic fields on galactic and extragalactic scales.

\begin{acknowledgments}
DM acknowledges the support of Banwarilal Bhalotia College, affiliated with Kazi Nazrul University, where this research was conducted during a summer research project. The authors appreciate the assistance of ChatGPT (OpenAI) in copy editing. AG acknowledges the Inter-University Centre for Astronomy and Astrophysics (IUCAA), Pune, for support through its Associateship Programme and for providing access to the computational facilities used in this work. We thank the reviewer for
the careful evaluation of our manuscript and for the constructive comments and
suggestions, which have helped us improve the paper.
\end{acknowledgments}

\begin{contribution}
DM, an MSc student, performed the data analysis and contributed to writing the initial draft. AG conceived the project, guided the methodology, and contributed to writing and finalising the manuscript. DB helped in preparing the final draft of the manuscript. \end{contribution}

\appendix

\section{Residuals of the fractional Stokes fit}
\label{app:res}

\begin{figure*}[!htbp]
    \centering
    \includegraphics[width=\textwidth]{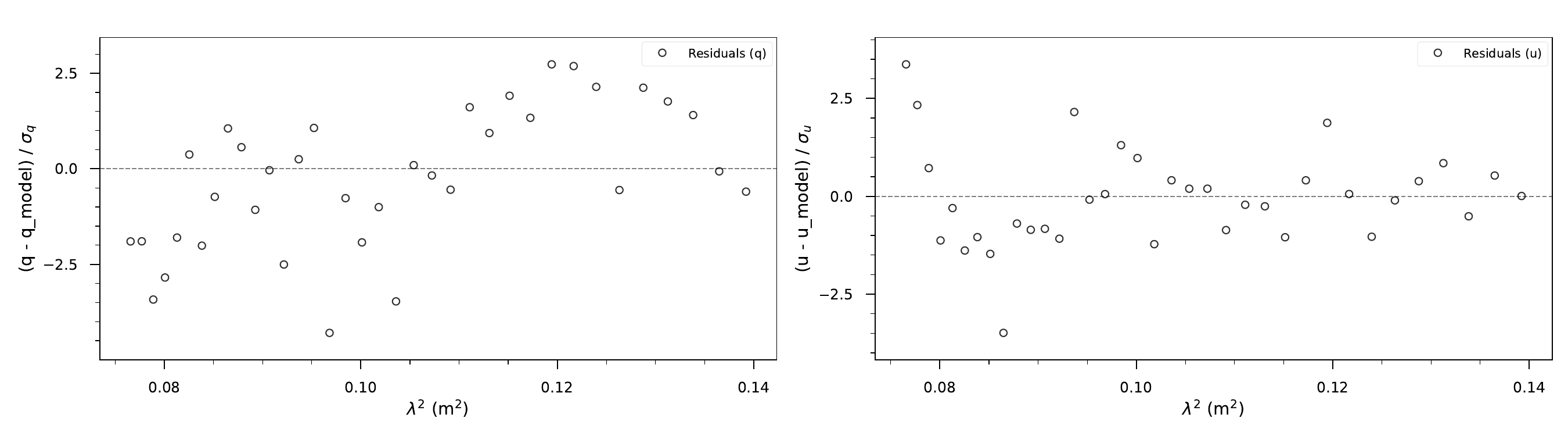}
    \caption{Residuals of the best-fitting \texttt{m5} model for the fractional
Stokes $q$ and $u$ spectra of RACS\_0900$-$28\_7036, normalised by the
measurement uncertainties $\sigma_q$ and $\sigma_u$.}
    \label{fig:qures}
\end{figure*}

Figure~\ref{fig:qures} shows the residuals between the observed fractional Stokes parameters $q$ and $u$ and the predictions of the best-fitting \texttt{m5} model. The residuals are broadly scattered around zero without prominent systematic trends, suggesting that the model adequately captures the primary Faraday structure of RACS\_0900$-$28\_7036.

However, the relatively higher reduced chi-squared value, $\chi^2_\mathrm{red} = 2.93$, suggests that the formal measurement uncertainties may be underestimated or that the model does not account for all subtle features in the polarization spectra. Some structured residuals remain at specific wavelengths, likely reflecting additional unresolved Faraday complexity, such as multiple unresolved synchrotron-emitting regions, gradients in Faraday depth across the source, or non-Gaussian turbulence in the foreground magneto-ionic medium \citep{Sokoloff_1998, OSullivan_2012, Anderson_2016}. These residuals suggest that, although the \texttt{m5} model captures the main features of the data, there may be additional small-scale magneto-ionic structure not fully resolved by the current model \citep{Green_2006, Hutschenreuter_2022}.

\bibliography{main}{}

@ARTICLE{AkahoriRyu2011,
       author = {{Akahori}, T. and {Ryu}, D.},
        title = "{Faraday Rotation Measure Due to the Intergalactic Magnetic Field}",
      journal = {The Astrophysical Journal},
         year = 2011,
        volume = {738},
        pages = {134},
          doi = {10.1088/0004-637X/738/2/134}
}

@ARTICLE{Anderson_2015,
       author = {{Anderson}, C.~S. and {Gaensler}, B.~M. and {Feain}, I.~J. and {Franzen}, T.~M.~O.},
        title = "{Broadband Radio Polarimetry and Faraday Rotation of 563 Extragalactic Radio Sources}",
      journal = {\apj},
         year = 2015,
        month = dec,
       volume = {815},
       number = {1},
          eid = {49},
        pages = {49},
          doi = {10.1088/0004-637X/815/1/49},
archivePrefix = {arXiv},
       eprint = {1511.04080},
 primaryClass = {astro-ph.GA},
       adsurl = {https://ui.adsabs.harvard.edu/abs/2015ApJ...815...49A}
}

@ARTICLE{Anderson_2016,
       author = {{Anderson}, C.~S. and {Gaensler}, B.~M. and {Feain}, I.~J.},
        title = "{A Study of Broadband Faraday Rotation and Polarization Behavior over 1.3--10 GHz in 36 Discrete Radio Sources}",
      journal = {\apj},
         year = 2016,
        month = jul,
       volume = {825},
       number = {1},
          eid = {59},
        pages = {59},
          doi = {10.3847/0004-637X/825/1/59},
archivePrefix = {arXiv},
       eprint = {1604.01403},
 primaryClass = {astro-ph.HE},
       adsurl = {https://ui.adsabs.harvard.edu/abs/2016ApJ...825...59A}
}

@article{Anderson_2021,
  author  = {Anderson, C. S. and Heald, G. H. and Eilek, J. A. and Lenc, E. and Gaensler, B. M. and Rudnick, Lawrence and Van Eck, C. L. and O'Sullivan, S. P. and Stil, J. M. and Chippendale, A. and et al.},
  title   = {Early Science from POSSUM: Shocks, turbulence, and a massive new reservoir of ionised gas in the Fornax cluster},
  journal = {Publications of the Astronomical Society of Australia},
  year    = {2021},
  volume  = {38},
  pages   = {e020},
  doi     = {10.1017/pasa.2021.14}
}

@article{Anderson_2023,
  author = {Anderson, C. S. and others},
  title = {{The Rapid ASKAP Continuum Survey III: Spectra and Polarisation in Cutouts of Extragalactic sources from RACS (SPICE-RACS) DR1}},
  journal = {Monthly Notices of the Royal Astronomical Society},
  year = {2023},
  eprint = {2307.07207},
  doi = {10.1093/mnras/stad2458}
}

@ARTICLE{Anderson_2024,
       author = {{Anderson}, C.~S. and {McClure-Griffiths}, N.~M. and {Rudnick}, L. and {Gaensler}, B.~M. and {O'Sullivan}, S.~P. and {Bradbury}, P. and {Akahori}, T. and {Baidoo}, L. and {Carretti}, E. and {Heald}, G. and {Kapi{\'n}ska}, A.~D. and {Lenc}, E. and {Leahy}, J.~P. and {Loi}, F. and {Riseley}, C.~J. and {Scaife}, A.~M.~M. and {Stil}, J.~M. and {Vacca}, V. and {Van Eck}, C.~L. and {West}, J.~L.},
        title = "{Probing the magnetized gas distribution in galaxy groups and the cosmic web with POSSUM Faraday rotation measures}",
      journal = {\mnras},
         year = 2024,
       volume = {533},
       number = {4},
        pages = {4068-4080},
          doi = {10.1093/mnras/stae1954},
archivePrefix = {arXiv},
       eprint = {2407.20325},
       adsurl = {https://ui.adsabs.harvard.edu/abs/2024MNRAS.533.4068A},
}

@ARTICLE{Brentjens_2005,
       author = {{Brentjens}, M.~A. and {de Bruyn}, A.~G.},
        title = "{Faraday rotation measure synthesis}",
      journal = {\aap},
         year = 2005,
        month = oct,
       volume = {441},
       number = {3},
        pages = {1217-1228},
          doi = {10.1051/0004-6361:20052990},
archivePrefix = {arXiv},
       eprint = {astro-ph/0507349},
 primaryClass = {astro-ph},
       adsurl = {https://ui.adsabs.harvard.edu/abs/2005A&A...441.1217B}
}

@article{Brentjens_2019,
  author = {Brentjens, M. A. and de Bruyn, A. G. and Pizzo, R. F. and Shimwell, T. W. and O'Sullivan, S. P. and others},
  title = {External Faraday dispersion in the LOFAR Two-metre Sky Survey},
  journal = {Astronomy \& Astrophysics},
  year = {2019},
  volume = {628},
  pages = {A10},
  doi = {10.1051/0004-6361/201935651},
  publisher = {EDP Sciences}
}

@ARTICLE{Bonafede2010,
       author = {{Bonafede}, A. and {Feretti}, L. and {Murgia}, M. and
                 {Govoni}, F. and {Giovannini}, G. and others},
        title = "{The Coma cluster magnetic field from Faraday rotation measures}",
      journal = {Astronomy \& Astrophysics},
         year = 2010,
        volume = {513},
        pages = {A30},
          doi = {10.1051/0004-6361/200913696}
}

@article{Burn_1966,
    author = {Burn, B. J.},
    title = {On the Depolarization of Discrete Radio Sources by Faraday Dispersion},
    journal = {Monthly Notices of the Royal Astronomical Society},
    volume = {133},
    number = {1},
    pages = {67-83},
    year = {1966},
    month = {07},
    issn = {0035-8711},
    doi = {10.1093/mnras/133.1.67},
    url = {https://doi.org/10.1093/mnras/133.1.67},
    eprint = {https://academic.oup.com/mnras/article-pdf/133/1/67/8078603/mnras133-0067.pdf},
}

@ARTICLE{Carretti_2022,
       author = {{Carretti}, E. and {Vacca}, V. and {O'Sullivan}, S.~P. and {Heald}, G.~H. and {Horellou}, C. and {R{\"o}ttgering}, H.~J.~A. and {Scaife}, A.~M.~M. and {Shimwell}, T.~W. and {Shulevski}, A. and {Stuardi}, C. and {Vernstrom}, T.},
        title = "{Magnetic field strength in cosmic web filaments}",
      journal = {\mnras},
         year = 2022,
        month = may,
       volume = {512},
       number = {1},
        pages = {945-3959},
          doi = {10.1093/mnras/stac384},
archivePrefix = {arXiv},
       eprint = {2202.04607},
 primaryClass = {astro-ph.CO},
       adsurl = {https://ui.adsabs.harvard.edu/abs/2022MNRAS.512..945C}
}

@ARTICLE{Condon98,
       author = {{Condon}, J.~J. and {Cotton}, W.~D. and {Greisen}, E.~W. and 
                 {Yin}, Q.~F. and {Perley}, R.~A. and {Taylor}, G.~B. and 
                 {Broderick}, J.~J.},
        title = "{The NRAO VLA Sky Survey}",
      journal = {The Astronomical Journal},
         year = 1998,
        volume = {115},
        pages = {1693--1716},
          doi = {10.1086/300337}
}

@ARTICLE{Dickey_2022,
       author = {{Dickey}, John M. and {West}, Jennifer and {Thomson}, Alec J.~M. and {Landecker}, T.~L. and {Bracco}, A. and {Carretti}, E. and {Han}, J.~L. and {Hill}, A.~S. and {Ma}, Y.~K. and {Mao}, S.~A.},
        title = "{Structure in the Magnetic Field of the Milky Way Disk and Halo Traced by Faraday Rotation}",
      journal = {\apj},
         year = 2022,
        month = nov,
       volume = {940},
       number = {1},
          eid = {75},
        pages = {75},
          doi = {10.3847/1538-4357/ac94ce},
archivePrefix = {arXiv},
       eprint = {2210.08112},
 primaryClass = {astro-ph.GA},
       adsurl = {https://harvard.edu}
}

@ARTICLE{Ferri_2021,
       author = {{Ferri{\`e}re}, K. and {West}, J.~L. and {Jaffe}, T.~R.},
        title = "{The correct sense of Faraday rotation}",
      journal = {\mnras},
         year = 2021,
        month = nov,
       volume = {507},
       number = {4},
        pages = {4968-4982},
          doi = {10.1093/mnras/stab1641},
archivePrefix = {arXiv},
       eprint = {2106.03074},
 primaryClass = {astro-ph.GA},
       adsurl = {https://ui.adsabs.harvard.edu/abs/2021MNRAS.507.4968F}
}

@article{Feroz_2008,
   title={Multimodal nested sampling: an efficient and robust alternative to Markov Chain Monte Carlo methods for astronomical data analyses: Multimodal nested sampling},
   volume={384},
   ISSN={1365-2966},
   url={http://dx.doi.org/10.1111/j.1365-2966.2007.12353.x},
   DOI={10.1111/j.1365-2966.2007.12353.x},
   number={2},
   journal={Monthly Notices of the Royal Astronomical Society},
   publisher={Oxford University Press (OUP)},
   author={Feroz, F. and Hobson, M. P.},
   year={2008},
   month=jan, pages={449–463} }

@article{Feroz_2009,
   title={MultiNest: an efficient and robust Bayesian inference tool for cosmology and particle physics},
   volume={398},
   ISSN={1365-2966},
   url={http://dx.doi.org/10.1111/j.1365-2966.2009.14548.x},
   DOI={10.1111/j.1365-2966.2009.14548.x},
   number={4},
   journal={Monthly Notices of the Royal Astronomical Society},
   publisher={Oxford University Press (OUP)},
   author={Feroz, F. and Hobson, M. P. and Bridges, M.},
   year={2009},
   month=oct, pages={1601–1614} }

@article{Feroz_2019,
   title={Importance Nested Sampling and the MultiNest Algorithm},
   volume={2},
   ISSN={2565-6120},
   url={http://dx.doi.org/10.21105/astro.1306.2144},
   DOI={10.21105/astro.1306.2144},
   number={1},
   journal={The Open Journal of Astrophysics},
   publisher={Maynooth University},
   author={Feroz, Farhan and Hobson, Michael P. and Cameron, Ewan and Pettitt, Anthony N.},
   year={2019},
   month=nov }

@article{Farnsworth_2011,
   title={INTEGRATED POLARIZATION OF SOURCES AT λ ∼ 1 m AND NEW ROTATION MEASURE AMBIGUITIES},
   volume={141},
   ISSN={1538-3881},
   url={http://dx.doi.org/10.1088/0004-6256/141/6/191},
   DOI={10.1088/0004-6256/141/6/191},
   number={6},
   journal={The Astronomical Journal},
   publisher={American Astronomical Society},
   author={Farnsworth, Damon and Rudnick, Lawrence and Brown, Shea},
   year={2011},
   month=may, pages={191} }

@article{Gaensler_2005,
  author = {Gaensler, B. M. and Beck, R. and Feretti, L.},
  title = {Magnetic Fields in the Interstellar Medium and in Galaxies},
  journal = {New Astronomy Reviews},
  year = {2005},
  volume = {49},
  pages = {227},
  doi = {10.1016/j.newar.2004.12.003}
}

@article{Gaensler_2004,
  author  = {Gaensler, B. M. and Beck, R. and Feretti, L.},
  title   = {The origin and evolution of cosmic magnetism},
  journal = {New Astronomy Reviews},
  year    = {2004},
  month   = dec,
  volume  = {48},
  number  = {11-12},
  pages   = {1003--1012},
  doi     = {10.1016/j.newar.2004.09.014}
}

@article{Gaensler_2025,
  title     = {The Polarisation Sky Survey of the Universe’s Magnetism ({POSSUM}): Science goals and survey description},
  volume    = {42},
  doi       = {10.1017/pasa.2025.10031},
  number    = {e091},
  journal   = {Publications of the Astronomical Society of Australia},
  publisher = {Cambridge University Press},
  author    = {Gaensler, B. M. and Heald, G. H. and McClure-Griffiths, N. M. and Anderson, C. S. and Van Eck, C. L. and West, J. L. and Thomson, A. J. M. and Leahy, J. P. and Rudnick, L. and Ma, Y. K. and et al.},
  year      = {2025},
  pages     = {e091}
}

@article{Gaensler_2025_CGM,
  author = {Malik, Sunil and O'Sullivan, S. P. and Thomson, A. J. M. and Anderson, C. S. and Van Eck, C. and Rudnick, L. and Seta, Amit and Gaensler, B. M. and Ma, Y. K. and Akahori, Takuya and Alonso-L{\'o}pez, D. and Br{\"u}ggen, M. and Carretti, E. and Duchesne, S. W. and Galvin, T. J. and Heald, G. and Hlinka, O. and Khadir, A. and Mao, S. A. and Omae, R.},
  title = {Magnetised CGM Gas at z$\sim$1 revealed by SPICE-RACS},
  journal = {arXiv e-prints},
  year = {2026},
  eprint = {2605.16924},
  archivePrefix = {arXiv},
  primaryClass = {astro-ph.GA},
  adsurl = {https://ui.adsabs.harvard.edu/abs/2026arXiv260516924M},
  note = {arXiv:2605.16924}
}

@article{Green_2006,
  author = {Green, J. A.},
  title = {Enhanced Small-Scale Faraday Rotation in the Galactic Spiral Arms},
  journal = {ApJ Letters},
  volume = {637},
  pages = {L33--L36},
  year = {2006},
  doi = {10.1086/500543}
}

@ARTICLE{Goodlet_2004,
       author = {{Goodlet}, J.~A. and {Kaiser}, C.~R. and {Best}, P.~N. and {Dennett-Thorpe}, J.},
        title = "{The depolarization properties of powerful radio sources: breaking the radio power versus redshift degeneracy}",
      journal = {\mnras},
         year = 2004,
        month = jan,
       volume = {347},
       number = {2},
        pages = {508-540},
          doi = {10.1111/j.1365-2966.2004.07225.x},
archivePrefix = {arXiv},
       eprint = {astro-ph/0309529},
 primaryClass = {astro-ph},
       adsurl = {https://ui.adsabs.harvard.edu/abs/2004MNRAS.347..508G}
}

@ARTICLE{Harvey_2011,
       author = {{Harvey-Smith}, L. and {Madsen}, G.~J. and {Gaensler}, B.~M.},
        title = "{Magnetic Fields in Large-diameter H II Regions Revealed by Faraday Rotation}",
      journal = {\apj},
         year = 2011,
        month = aug,
       volume = {736},
       number = {2},
          eid = {83},
        pages = {83},
          doi = {10.1088/0004-637X/736/2/83},
archivePrefix = {arXiv},
       eprint = {1106.0931},
 primaryClass = {astro-ph.GA},
       adsurl = {https://ui.adsabs.harvard.edu/abs/2011ApJ...736...83H}
}

@ARTICLE{Hotan_2021,
       author = {{Hotan}, A.~W. and {Bunton}, J.~D. and {Chippendale}, A.~P. and {Whiting}, M. and {Amy}, S.~W. and {Bane}, M.~R. and {Barker}, S. and {Barrett}, J. and {Batty}, M. and {Bell}, M.~E. and {Bock}, D. C.~J. and {Bolton}, R. and {Brodrick}, D. and {Burgess}, A. and {Busch}, M. and {Carretti}, E. and {Chekkala}, R. and {Collier}, J.~D. and {Cooray}, F.~R. and {Cornwell}, T.~J. and {De Boer}, B. and {Diamond}, P.~J. and {Edwards}, P.~G. and {Ekers}, R.~D. and {Feain}, I. and {Ferris}, R.~H. and {Forsyth}, R. and {Gough}, R.~G. and {Grancea}, A. and {Gupta}, N. and {Guzman}, J.~C. and {Hampson}, G.~A. and {Harvey-Smith}, L. and {Hay}, S.~G. and {Hayman}, D.~B. and {Heald}, G. and {Hicks}, G. and {Hobbs}, G. and {Hoepper}, S. and {Hood}, B.~C. and {Hopkins}, A.~M. and {Humphreys}, B. and {Indermuehle}, B. and {Jacka}, C.~E. and {Jackson}, C.~A. and {Jackson}, S. and {Jeganathan}, K. and {Johnston}, S. and {Joseph}, J. and {Kamphuis}, P. and {Karastergiou}, A. and {Kesteven}, M.~J. and {Kiryaly}, Z. and {Koribalski}, B.~S. and {Leach}, M. and {Lenc}, E. and {Lensson}, E. and {Li}, L. and {Lynch}, C.~R. and {MacLeod}, M. and {Marquarding}, M. and {McClure-Griffiths}, N.~M. and {McConnell}, D. and {Mirtschin}, P. and {O'Sullivan}, S.~P. and {Pearce}, S. and {Pekal}, R. and {Phillips}, C.~J. and {Price}, D.~C. and {Rezaei}, M. and {Riseley}, C.~J. and {Roberts}, P. and {Sault}, R.~J. and {Schinckel}, A.~E.~T. and {Scott}, P. and {Severs}, S.~R. and {Shannon}, R.~M. and {Shields}, M. and {Shimwell}, T.~W. and {Stil}, J.~M. and {Sweetnam}, T.~W. and {Tzioumis}, A.~K. and {Voronkov}, M.~A. and {Wall}, N. and {West}, J. and {Wolleben}, M.},
        title = "{Australian Square Kilometre Array Pathfinder}",
      journal = {Publications of the Astronomical Society of Australia},
         year = 2021,
        month = mar,
       volume = {38},
          eid = {e009},
        pages = {e009},
          doi = {10.1017/pasa.2021.1},
archivePrefix = {arXiv},
       eprint = {2102.01870},
 primaryClass = {astro-ph.IM},
       adsurl = {https://harvard.edu}
}

@ARTICLE{Hutschenreuter22,
       author = {{Hutschenreuter}, S. and {Anderson}, C.~S. and {Betti}, S. and {Bower}, G.~C. and {Brown}, J.-A. and {Br{\"u}ggen}, M. and {Carretti}, E. and {Clarke}, T. and {Clegg}, A. and {Costa}, A. and {Croft}, S. and {Van Eck}, C. and {Gaensler}, B.~M. and {de Gasperin}, F. and {Haverkorn}, M. and {Heald}, G. and {Hull}, C.~L.~H. and {Inoue}, M. and {Johnston-Hollitt}, M. and {Kaczmarek}, J. and {Law}, C. and {Ma}, Y.~K. and {MacMahon}, D. and {Mao}, S.~A. and {Riseley}, C. and {Roy}, S. and {Shanahan}, R. and {Shimwell}, T. and {Stil}, J. and {Sobey}, C. and {O'Sullivan}, S.~P. and {Tasse}, C. and {Vacca}, V. and {Vernstrom}, T. and {Williams}, P.~K.~G. and {Wright}, M. and {En{\ss}lin}, T.~A.},
        title = "{The Galactic Faraday rotation sky 2020}",
      journal = {\aap},
         year = 2022,
        month = jan,
       volume = {657},
          eid = {A43},
        pages = {A43},
          doi = {10.1051/0004-6361/202140486},
archivePrefix = {arXiv},
       eprint = {2102.01709},
 primaryClass = {astro-ph.GA},
       adsurl = {https://ui.adsabs.harvard.edu/abs/2022A&A...657A..43H}
}

@article{Hutschenreuter_2022,
  author = {Hutschenreuter, S. et al.},
  title = {Full-Sky Polarimetric Radio Continuum Reconstruction at 1.4 GHz},
  journal = {A\&A},
  volume = {663},
  pages = {A127},
  year = {2022},
  doi = {10.1051/0004-6361/202142733}
}

@ARTICLE{Intema17,
       author = {{Intema}, H.~T. and {Jagannathan}, P. and {Mooley}, K.~P. and 
                 {Frail}, D.~A.},
        title = "{The GMRT 150 MHz all-sky radio survey: First alternative data release TGSS ADR1}",
      journal = {Astronomy \& Astrophysics},
         year = 2017,
        volume = {598},
        pages = {A78},
          doi = {10.1051/0004-6361/201628536}
}

@ARTICLE{Jelic_2014,
       author = {{Jeli{\'c}}, V. and {de Bruyn}, A.~G. and {Mevius}, M. and {Abdalla}, F.~B. and {Asad}, K.~M.~B. and {Bernardi}, G. and {Brentjens}, M.~A. and {Bus}, S. and {Chapman}, E. and {Ciardi}, B. and {Daiboo}, S. and {Fernandez}, E.~R. and {Ghosh}, A. and {Harker}, G. and {Jensen}, H. and {Kazemi}, S. and {Koopmans}, L.~V.~E. and {Labropoulos}, P. and {Martinez-Rubi}, O. and {Mellema}, G. and {Offringa}, A.~R. and {Pandey}, V.~N. and {Patil}, A.~H. and {Thomas}, R.~M. and {Vedantham}, H.~K. and {Veligatla}, V. and {Yatawatta}, S. and {Zaroubi}, S. and {Alexov}, A. and {Anderson}, J. and {Avruch}, I.~M. and {Beck}, R. and {Bell}, M.~E. and {Bentum}, M.~J. and {Best}, P. and {Bonafede}, A. and {Bregman}, J. and {Breitling}, F. and {Broderick}, J. and {Brouw}, W.~N. and {Br{\"u}ggen}, M. and {Butcher}, H.~R. and {Conway}, J.~E. and {de Gasperin}, F. and {de Geus}, E. and {Deller}, A. and {Dettmar}, R. -J. and {Duscha}, S. and {Eisl{\"o}ffel}, J. and {Engels}, D. and {Falcke}, H. and {Fallows}, R.~A. and {Fender}, R. and {Ferrari}, C. and {Frieswijk}, W. and {Garrett}, M.~A. and {Grie{\ss}meier}, J. and {Gunst}, A.~W. and {Hamaker}, J.~P. and {Hassall}, T.~E. and {Haverkorn}, M. and {Heald}, G. and {Hessels}, J.~W.~T. and {Hoeft}, M. and {H{\"o}randel}, J. and {Horneffer}, A. and {van der Horst}, A. and {Iacobelli}, M. and {Juette}, E. and {Karastergiou}, A. and {Kondratiev}, V.~I. and {Kramer}, M. and {Kuniyoshi}, M. and {Kuper}, G. and {van Leeuwen}, J. and {Maat}, P. and {Mann}, G. and {McKay-Bukowski}, D. and {McKean}, J.~P. and {Munk}, H. and {Nelles}, A. and {Norden}, M.~J. and {Paas}, H. and {Pandey-Pommier}, M. and {Pietka}, G. and {Pizzo}, R. and {Polatidis}, A.~G. and {Reich}, W. and {R{\"o}ttgering}, H. and {Rowlinson}, A. and {Scaife}, A.~M.~M. and {Schwarz}, D. and {Serylak}, M. and {Smirnov}, O. and {Steinmetz}, M. and {Stewart}, A. and {Tagger}, M. and {Tang}, Y. and {Tasse}, C. and {ter Veen}, S. and {Thoudam}, S. and {Toribio}, C. and {Vermeulen}, R. and {Vocks}, C. and {van Weeren}, R.~J. and {Wijers}, R.~A.~M.~J. and {Wijnholds}, S.~J. and {Wucknitz}, O. and {Zarka}, P.},
        title = "{Initial LOFAR observations of epoch of reionization windows. II. Diffuse polarized emission in the ELAIS-N1 field}",
      journal = {\aap},
         year = 2014,
        month = aug,
       volume = {568},
          eid = {A101},
        pages = {A101},
          doi = {10.1051/0004-6361/201423998},
archivePrefix = {arXiv},
       eprint = {1407.2093},
 primaryClass = {astro-ph.GA},
       adsurl = {https://ui.adsabs.harvard.edu/abs/2014A&A...568A.101J}
}

@article{Kass_1995,
author = {Robert E. Kass and Adrian E. Raftery},
title = {Bayes Factors},
journal = {Journal of the American Statistical Association},
volume = {90},
number = {430},
pages = {773--795},
year = {1995},
publisher = {ASA Website},
doi = {10.1080/01621459.1995.10476572},
URL = {  
        https://www.tandfonline.com/doi/abs/10.1080/01621459.1995.10476572
  
},
eprint = { 
        https://www.tandfonline.com/doi/pdf/10.1080/01621459.1995.10476572
 
}
}

@ARTICLE{Laing_2008,
       author = {{Laing}, R.~A. and {Bridle}, A.~H. and {Parma}, P. and {Murgia}, M.},
        title = "{Structures of the magnetoionic media around the Fanaroff-Riley Class I radio galaxies 3C31 and Hydra A}",
      journal = {\mnras},
         year = 2008,
        month = dec,
       volume = {391},
       number = {2},
        pages = {521-549},
          doi = {10.1111/j.1365-2966.2008.13895.x},
archivePrefix = {arXiv},
       eprint = {0809.2411},
 primaryClass = {astro-ph},
       adsurl = {https://ui.adsabs.harvard.edu/abs/2008MNRAS.391..521L}
}

@ARTICLE{Livingston_2022,
       author = {{Livingston}, J.~D. and {McClure-Griffiths}, N.~M. and {Mao}, S.~A. and {Gaensler}, B.~M. and {Heald}, G. and {Lenc}, E. and {Dickey}, J.~M.},
        title = "{A radio polarization study of magnetic fields in the Small Magellanic Cloud}",
      journal = {\mnras},
         year = 2022,
        month = feb,
       volume = {510},
       number = {1},
        pages = {260-275},
          doi = {10.1093/mnras/stab3375},
archivePrefix = {arXiv},
       eprint = {2112.04044},
 primaryClass = {astro-ph.GA},
       adsurl = {https://harvard.edu}
}

@article{LOFAR_2023,
   title={The Faraday Rotation Measure Grid of the LOFAR Two-metre Sky Survey: Data Release 2},
   volume={519},
   ISSN={1365-2966},
   url={http://dx.doi.org/10.1093/mnras/stac3820},
   DOI={10.1093/mnras/stac3820},
   number={4},
   journal={Monthly Notices of the Royal Astronomical Society},
   publisher={Oxford University Press (OUP)},
   author={O’Sullivan, S P and Shimwell, T W and Hardcastle, M J and Tasse, C and Heald, G and Carretti, E and Brüggen, M and Vacca, V and Sobey, C and VanEck, C L and Horellou, C and Beck, R and Bilicki, M and Bourke, S and Botteon, A and Croston, J H and Drabent, A and Duncan, K and Heesen, V and Ideguchi, S and Kirwan, M and Lawlor, L and Mingo, B and Nikiel-Wroczyński, B and Piotrowska, J and Scaife, A M M and van-Weeren, R J},
   year={2023},
   month=jan, pages={5723–5742} 
}

@ARTICLE{Loi25,
       author = {{Loi}, F. and {Serra}, P. and {Murgia}, M. and {Govoni}, F. and {Vacca}, V. and {Maccagni}, F. and {Kleiner}, D. and {Kamphuis}, P.},
        title = "{The MeerKAT Fornax Survey: IV. A close look at the cluster physics through the densest rotation measure grid}",
      journal = {\aap},
         year = 2025,
        month = feb,
       volume = {694},
          eid = {A125},
        pages = {A125},
          doi = {10.1051/0004-6361/202451711},
archivePrefix = {arXiv},
       eprint = {2501.05519},
 primaryClass = {astro-ph.CO},
       adsurl = {https://ui.adsabs.harvard.edu/abs/2025A&A...694A.125L}
}

@ARTICLE{Mao_2010,
       author = {{Mao}, S.~A. and {Gaensler}, B.~M. and {Haverkorn}, M. and {Zweibel}, E.~G. and {Madsen}, G.~J. and {McClure-Griffiths}, N.~M. and {Shukurov}, A. and {Kronberg}, P.~P.},
        title = "{A Survey of Extragalactic Faraday Rotation at High Galactic Latitude: The Vertical Magnetic Field of the Milky Way towards the Galactic Poles}",
      journal = {\apj},
         year = 2010,
        month = may,
       volume = {714},
       number = {2},
        pages = {1170-1186},
          doi = {10.1088/0004-637X/714/2/1170},
archivePrefix = {arXiv},
       eprint = {1003.4519},
 primaryClass = {astro-ph.GA},
       adsurl = {https://harvard.edu}
}

@ARTICLE{McConnell_2020,
       author = {{McConnell}, D. and {Hale}, C.~L. and {Lenc}, E. and {Banfield}, J.~K. and others},
        title = "{The Rapid ASKAP Continuum Survey I: Design and first results}",
      journal = {Publications of the Astronomical Society of Australia},
         year = 2020,
       volume = {37},
          eid = {e048},
          doi = {10.1017/pasa.2020.41},
       adsurl = {https://harvard.edu},
}

@ARTICLE{Merloni2024,
       author = {{Merloni}, A. and {Lamer}, G. and {Liu}, T. and 
                 {Ramos-Ceja}, M.~E. and {Brunner}, H. and 
                 {Bulbul}, E. and {Dennerl}, K. and {Doroshenko}, V. and
                 {Freyberg}, M.~J. and {Friedrich}, S. and others},
        title = "{The SRG/eROSITA all-sky survey. First X-ray catalogues and data release of the western Galactic hemisphere}",
      journal = {Astronomy \& Astrophysics},
         year = 2024,
        volume = {682},
        pages = {A34},
          doi = {10.1051/0004-6361/202347165}
}

@misc{Mevius_2018,
       author = {{Mevius}, Maaijke},
        title = "{RMextract: Ionospheric Faraday Rotation calculator}",
 howpublished = {Astrophysics Source Code Library, record ascl:1806.024},
         year = 2018,
        month = jun,
          eid = {ascl:1806.024},
       adsurl = {https://ui.adsabs.harvard.edu/abs/2018ascl.soft06024M}
}

@MISC{NED,
       author = {{NASA/IPAC Extragalactic Database (NED)}},
        title = "{NASA/IPAC Extragalactic Database}",
         year = 2024,
 howpublished = {\url{https://ned.ipac.caltech.edu/}}
}

@ARTICLE{Oei_2023,
       author = {{Oei}, Martijn S.~S.~L. and {van Weeren}, Reinout J. and {Gast}, Aivin R.~D.~J.~G.~I.~B. and {Botteon}, Andrea and {Hardcastle}, Martin J. and {Dabhade}, Pratik and {Shimwell}, Tim W. and {R{\"o}ttgering}, Huub J.~A. and {Drabent}, Alexander},
        title = "{Measuring the giant radio galaxy length distribution with the LoTSS}",
      journal = {\aap},
         year = 2023,
        month = apr,
       volume = {672},
          eid = {A163},
        pages = {A163},
          doi = {10.1051/0004-6361/202243572},
archivePrefix = {arXiv},
       eprint = {2210.10234},
 primaryClass = {astro-ph.GA},
       adsurl = {https://ui.adsabs.harvard.edu/abs/2023A&A...672A.163O}
}

@article{OSullivan_2012,
    author = {O’Sullivan, S. P. and Brown, S. and Robishaw, T. and Schnitzeler, D. H. F. M. and McClure‐Griffiths, N. M. and Feain, I. J. and Taylor, A. R. and Gaensler, B. M. and Landecker, T. L. and Harvey‐Smith, L. and Carretti, E.},
    title = {Complex Faraday depth structure of active galactic nuclei as revealed by broad‐band radio polarimetry},
    journal = {Monthly Notices of the Royal Astronomical Society},
    volume = {421},
    number = {4},
    pages = {3300-3315},
    year = {2012},
    month = {04},
    issn = {0035-8711},
    doi = {10.1111/j.1365-2966.2012.20554.x},
    url = {https://doi.org/10.1111/j.1365-2966.2012.20554.x},
    eprint = {https://academic.oup.com/mnras/article-pdf/421/4/3300/3821486/mnras0421-3300.pdf},
}

@article{OSullivan_2015,
	doi = {10.1088/0004-637x/806/1/83},
	url = {https://doi.org},
	year = 2015,
	month = {jun},
	publisher = {IOP Publishing},
	volume = {806},
	number = {1},
	pages = {83},
	author = {S. P. O'Sullivan and B. M. Gaensler and M. A. Lara-L{\'o}pez and S. van Velzen and J. K. Banfield and J. S. Farnes},
	title = {THE MAGNETIC FIELD AND POLARIZATION PROPERTIES OF RADIO GALAXIES IN DIFFERENT ACCRETION STATES},
	journal = {The Astrophysical Journal},
	adsurl = {https://harvard.edu}
}

@article{OSullivan_2017,
    author = "O'Sullivan, S. P. and Purcell, C. R. and Anderson, C. S. and Farnes, J. S. and Sun, X. H. and Gaensler, B. M.",
    title = "{Broad-band, radio spectro-polarimetric study of 100 radiative-mode and jet-mode AGN}",
    eprint = "1705.00102",
    archivePrefix = "arXiv",
    primaryClass = "astro-ph.GA",
    doi = "10.1093/mnras/stx1133",
    journal = "Mon. Not. Roy. Astron. Soc.",
    volume = "469",
    number = "4",
    pages = "4034--4062",
    year = "2017"
}

@article{OSullivan_2018,
  author = {Anderson, C. S. and Gaensler, B. M. and Heald, G. H. and O'Sullivan, S. P. and Kaczmarek, J. F. and Feain, I. J.},
  title = {Broadband radio polarimetry of Fornax A, I: Depolarized patches generated by advected thermal material from NGC 1316},
  journal = {Monthly Notices of the Royal Astronomical Society},
  year = {2018},
  volume = {479},
  number = {2},
  pages = {1436--1450},
  doi = {10.1093/mnras/sty1747},
  eprint = {1802.04812},
  archivePrefix = {arXiv},
  primaryClass = {astro-ph.GA}
}

@Article{Pasetto_2016,
AUTHOR = {Pasetto, Alice and Carrasco-González, Carlos and Bruni, Gabriele and Basu, Aritra and O’Sullivan, Shane and Kraus, Alex and Mack, Karl-Heinz},
TITLE = {JVLA Wideband Polarimetry Observations on a Sample of High Rotation Measure Sources},
JOURNAL = {Galaxies},
VOLUME = {4},
YEAR = {2016},
NUMBER = {4},
ARTICLE-NUMBER = {66},
URL = {https://www.mdpi.com/2075-4434/4/4/66},
ISSN = {2075-4434},
DOI = {10.3390/galaxies4040066}
}

@ARTICLE{Pasetto_2018,
       author = {{Pasetto}, Alice and {Carrasco-Gonz{\'a}lez}, Carlos and {O'Sullivan}, Shane and {Basu}, Aritra and {Bruni}, Gabriele and {Kraus}, Alex and {Curiel}, Salvador and {Mack}, Karl-Heinz},
        title = "{Broadband radio spectro-polarimetric observations of high-Faraday-rotation-measure AGN}",
      journal = {\aap},
         year = 2018,
        month = jun,
       volume = {613},
          eid = {A74},
        pages = {A74},
          doi = {10.1051/0004-6361/201731804},
archivePrefix = {arXiv},
       eprint = {1801.09731},
 primaryClass = {astro-ph.GA},
       adsurl = {https://ui.adsabs.harvard.edu/abs/2018A&A...613A..74P}
}

@Article{Passeto_2021,
AUTHOR = {Pasetto, Alice},
TITLE = {Message in a Bottle: Unveiling the Magneto-Ionic Complexity of AGNs through the Stokes QU-Fitting Technique},
JOURNAL = {Galaxies},
VOLUME = {9},
YEAR = {2021},
NUMBER = {3},
ARTICLE-NUMBER = {56},
URL = {https://www.mdpi.com/2075-4434/9/3/56},
ISSN = {2075-4434},
DOI = {10.3390/galaxies9030056}
}

@ARTICLE{Paul_2026,
       author = {{Paul}, Samantha Sneha and {Ghosh}, Abhik},
        title = "{Faraday Depolarization Study of a Radio Galaxy Using LOFAR Two-metre Sky Survey: Data Release 2}",
      journal = {The Open Journal of Astrophysics},
         year = 2026,
        month = feb,
       volume = {9},
        pages = {57500},
          doi = {10.33232/001c.157500},
archivePrefix = {arXiv},
       eprint = {2510.00440},
 primaryClass = {astro-ph.CO},
       adsurl = {https://ui.adsabs.harvard.edu/abs/2026OJAp....957500P}
}

@phdthesis{Pirasthesis_2024,
  author    = {Sara Piras},
  title     = {Polarization in the ELAIS-N1 LOFAR Deep Field},
  school    = {Chalmers University of Technology, Department of Space, Earth, and Environment},
  address   = {Gothenburg, Sweden},
  year      = {2024},
  type      = {PhD thesis},
  url       = {https://research.chalmers.se/publication/543484/file/543484_Fulltext.pdf},
  isbn      = {978-91-8103-129-4},
  note      = {Doktorsavhandlingar vid Chalmers tekniska högskola, Ny series nr 5587}
}

@misc{Purcell_2020,
       author = {{Purcell}, C.~R. and {Van Eck}, C.~L. and {West}, J. and {Sun}, X.~H. and {Gaensler}, B.~M.},
        title = "{RM-Tools: Rotation measure (RM) synthesis and Stokes QU-fitting}",
 howpublished = {Astrophysics Source Code Library, record ascl:2005.003},
         year = 2020,
        month = may,
          eid = {ascl:2005.003},
       adsurl = {https://ui.adsabs.harvard.edu/abs/2020ascl.soft05003P}
}

@article{Rybicki_Lightman_1979,
  author = {Rybicki, George B. and Lightman, Alan P.},
  title = {Radiative Processes in Astrophysics},
  journal = {Wiley-Interscience},
  year = {1979},
  pages = {394}
}

@ARTICLE{Rossetti_2008,
       author = {{Rossetti}, A. and {Dallacasa}, D. and {Fanti}, C. and
                 {Fanti}, R. and {Mack}, K.-H.},
        title = "{Faraday rotation and depolarization in compact radio sources}",
      journal = {\aap},
         year = 2008,
        volume = {487},
        pages = {865-872},
          doi = {10.1051/0004-6361:200809571},
          adsurl = {https://ui.adsabs.harvard.edu/abs/2008A%26A...487..865R},
}

@ARTICLE{Shimwell_2022,
       author = {{Shimwell}, T.~W. and {Hardcastle}, M.~J. and {Tasse}, C. and {Best}, P.~N. and {R{\"o}ttgering}, H.~J.~A. and {Williams}, W.~L. and {Botteon}, A. and {Drabent}, A. and {Mechev}, A. and {Shulevski}, A. and {van Weeren}, R.~J. and {Bester}, L. and {Br{\"u}ggen}, M. and {Brunetti}, G. and {Callingham}, J.~R. and {Chy{\.z}y}, K.~T. and {Conway}, J.~E. and {Dijkema}, T.~J. and {Duncan}, K. and {de Gasperin}, F. and {Hale}, C.~L. and {Haverkorn}, M. and {Hugo}, B. and {Jackson}, N. and {Mevius}, M. and {Miley}, G.~K. and {Morabito}, L.~K. and {Morganti}, R. and {Offringa}, A. and {Oonk}, J.~B.~R. and {Rafferty}, D. and {Sabater}, J. and {Smith}, D.~J.~B. and {Schwarz}, D.~J. and {Smirnov}, O. and {O'Sullivan}, S.~P. and {Vedantham}, H. and {White}, G.~J. and {Albert}, J.~G. and {Alegre}, L. and {Asabere}, B. and {Bacon}, D.~J. and {Bonafede}, A. and {Bonnassieux}, E. and {Brienza}, M. and {Bilicki}, M. and {Bonato}, M. and {Calistro Rivera}, G. and {Cassano}, R. and {Cochrane}, R. and {Croston}, J.~H. and {Cuciti}, V. and {Dallacasa}, D. and {Danezi}, A. and {Dettmar}, R.~J. and {Di Gennaro}, G. and {Edler}, H.~W. and {En{\ss}lin}, T.~A. and {Emig}, K.~L. and {Franzen}, T.~M.~O. and {Garc{\'\i}a-Vergara}, C. and {Grange}, Y.~G. and {G{\"u}rkan}, G. and {Hajduk}, M. and {Heald}, G. and {Heesen}, V. and {Hoang}, D.~N. and {Hoeft}, M. and {Horellou}, C. and {Iacobelli}, M. and {Jamrozy}, M. and {Jeli{\'c}}, V. and {Kondapally}, R. and {Kukreti}, P. and {Kunert-Bajraszewska}, M. and {Magliocchetti}, M. and {Mahatma}, V. and {Ma{\l}ek}, K. and {Mandal}, S. and {Massaro}, F. and {Meyer-Zhao}, Z. and {Mingo}, B. and {Mostert}, R.~I.~J. and {Nair}, D.~G. and {Nakoneczny}, S.~J. and {Nikiel-Wroczy{\'n}ski}, B. and {Orr{\'u}}, E. and {Pajdosz-{\'S}mierciak}, U. and {Pasini}, T. and {Prandoni}, I. and {van Piggelen}, H.~E. and {Rajpurohit}, K. and {Retana-Montenegro}, E. and {Riseley}, C.~J. and {Rowlinson}, A. and {Saxena}, A. and {Schrijvers}, C. and {Sweijen}, F. and {Siewert}, T.~M. and {Timmerman}, R. and {Vaccari}, M. and {Vink}, J. and {West}, J.~L. and {Wo{\l}owska}, A. and {Zhang}, X. and {Zheng}, J.},
        title = "{The LOFAR Two-metre Sky Survey. V. Second data release}",
      journal = {\aap},
         year = 2022,
        month = mar,
       volume = {659},
          eid = {A1},
        pages = {A1},
          doi = {10.1051/0004-6361/202142484},
archivePrefix = {arXiv},
       eprint = {2202.11733},
 primaryClass = {astro-ph.GA},
       adsurl = {https://ui.adsabs.harvard.edu/abs/2022A&A...659A...1S}
}

@ARTICLE{Shimwell_2026,
       author = {{Shimwell}, T.~W. and {Hardcastle}, M.~J. and {Tasse}, C. and {Drabent}, A. and {Botteon}, A. and {Williams}, W.~L. and {Best}, P.~N. and {R{\"o}ttgering}, H.~J.~A. and {Br{\"u}ggen}, M. and {Brunetti}, G. and {Callingham}, J.~R. and {Chy{\.z}y}, K.~T. and {Conway}, J.~E. and {De Gasperin}, F. and {Haverkorn}, M. and {Horellou}, C. and {Jackson}, N. and {Miley}, G.~K. and {Morabito}, L.~K. and {Morganti}, R. and {O'Sullivan}, S.~P. and {Schwarz}, D.~J. and {Smith}, D.~J.~B. and {van Weeren}, R.~J. and {Vedantham}, H.~K. and {White}, G.~J. and {Ahmadi}, A. and {Alegre}, L. and {Arias}, M. and {Asabere}, B. and {Bahr-Kalus}, B. and {Barkus}, B. and {Bilicki}, M. and {B{\"o}hme}, L. and {Brentjens}, M. and {Brienza}, M. and {Bomans}, D.~J. and {Bonafede}, A. and {Bonato}, M. and {Bonnassieux}, E. and {Boxelaar}, J.~M. and {Camera}, S. and {Cassano}, R. and {Chilufya}, J. and {Cianfaglione}, M. and {Croston}, J.~H. and {Cuciti}, V. and {Dabhade}, P. and {De Rubeis}, E. and {de Jong}, J.~M.~G.~H.~J. and {Dallacasa}, D. and {Dettmar}, R.~J. and {Duncan}, K.~J. and {Di Gennaro}, G. and {Edler}, H.~W. and {Groeneveld}, C. and {G{\"u}rkan}, G. and {Hajduk}, M. and {Hale}, C.~L. and {Heesen}, V. and {Hoang}, D.~N. and {Hoeft}, M. and {Holties}, H. and {Horton}, M.~A. and {Iacobelli}, M. and {Jamrozy}, M. and {Jarvis}, M.~J. and {Jelic}, V. and {Kadler}, M. and {Kondapally}, R. and {Kunert-Bajraszewska}, M. and {Loose}, M. and {Magliocchetti}, M. and {Ma{\l}ek}, K. and {Manzano}, C. and {McKean}, J.~P. and {Mevius}, M. and {Mingo}, B. and {Miskolczi}, A. and {Misra}, A. and {Mold{\'o}n}, J. and {Nair}, D.~G. and {Nakoneczny}, S.~J. and {Orru}, E. and {Pashapour-Ahmadabadi}, M. and {Pasini}, T. and {Petley}, J. and {Pierce}, J.~C.~S. and {Prandoni}, I. and {Rafferty}, D. and {Rajpurohit}, K. and {Riseley}, C.~J. and {Roberts}, I.~D. and {Sethi}, S. and {Shulevski}, A. and {Stein}, M. and {Stuardi}, C. and {Sweijen}, F. and {ter Veen}, S. and {Timmerman}, R. and {Vaccari}, M. and {Wijnholds}, S.},
        title = "{The LOFAR Two-metre Sky Survey: VII. Third Data Release}",
      journal = {\aap},
         year = 2026,
        month = mar,
       volume = {707},
          eid = {A198},
        pages = {A198},
          doi = {10.1051/0004-6361/202557749},
archivePrefix = {arXiv},
       eprint = {2602.15949},
 primaryClass = {astro-ph.GA},
       adsurl = {https://ui.adsabs.harvard.edu/abs/2026A&A...707A.198S}
}

@ARTICLE{Sinha_2026,
       author = {{Sinha}, Rudra and {Ghosh}, Abhik},
        title = "{Faraday Complexity and Depolarization in LOFAR Two-metre Sky Survey (LoTSS-DR2) Polarized Radio Sources}",
      journal = {arXiv e-prints},
         year = 2026,
        month = may,
          eid = {arXiv:2605.19226},
        pages = {arXiv:2605.19226},
          doi = {10.48550/arXiv.2605.19226},
archivePrefix = {arXiv},
       eprint = {2605.19226},
 primaryClass = {astro-ph.CO},
       adsurl = {https://ui.adsabs.harvard.edu/abs/2026arXiv260519226S}
}

@ARTICLE{Sokoloff_1998,
       author = {{Sokoloff}, D.~D. and {Bykov}, A.~A. and {Shukurov}, A. and {Berkhuijsen}, E.~M. and {Beck}, R. and {Poezd}, A.~D.},
        title = "{Depolarization and Faraday effects in galaxies}",
      journal = {\mnras},
         year = 1998,
        month = aug,
       volume = {299},
       number = {1},
        pages = {189-206},
          doi = {10.1046/j.1365-8711.1998.01782.x},
       adsurl = {https://ui.adsabs.harvard.edu/abs/1998MNRAS.299..189S}
}

@article{Sotomayor_2008,
  author = {Sotomayor-Beltran, C. and de Bruyn, A. G.},
  title = {Beam Depolarization in Interferometric Observations},
  journal = {A\&A},
  volume = {489},
  pages = {1013--1022},
  year = {2008}
}

@article{Sun_2015,
  author  = {Sun, X. H. and Rudnick, L. and Akahori, Takuya and Anderson, C. S. and Bell, M. R. and Bray, J. D. and Farnes, J. S. and Ideguchi, S. and Kumazaki, K. and O'Brien, T. and O'Sullivan, S. P. and Scaife, A. M. M. and Stepanov, R. and Stil, J. and Takahashi, K. and van Weeren, R. J. and Wolleben, M.},
  title   = {Comparison of Algorithms for Determination of Rotation Measure and Faraday Structure. I. 1100-1400 MHz},
  journal = {The Astronomical Journal},
  year    = {2015},
  volume  = {149},
  number  = {2},
  pages   = {60},
  doi     = {10.1088/0004-6256/149/2/60}
}

@article{Sun_2020,
  author        = {Sun, X. H. and Rudnick, L. and Akahori, T. and Wolleben, M. and Bell, M. R. and Bray, J. D. and Scaife, A. M. M. and others},
  title         = {Bayesian {RM} synthesis and {QU} fitting for complex {Faraday} structures},
  journal       = {The Astronomical Journal},
  volume        = {149},
  number        = {2},
  pages         = {60},
  year          = {2015},
  doi           = {10.1088/0004-6256/149/2/60},
  publisher     = {IOP Publishing}
}

@ARTICLE{Sun_2025,
       author = {{Sun}, X. and {Haverkorn}, M. and {Carretti}, E. and {Landecker}, T. and {Gaensler}, B.~M. and {Poppi}, S. and {Staveley-Smith}, L. and {Gao}, X. and {Han}, J.},
        title = "{The Southern Twenty-centimetre All-sky Polarization Survey (STAPS): Survey description and maps}",
      journal = {\aap},
         year = 2025,
        month = feb,
       volume = {694},
          eid = {A169},
        pages = {A169},
          doi = {10.1051/0004-6361/202453326},
archivePrefix = {arXiv},
       eprint = {2501.14203},
 primaryClass = {astro-ph.GA},
       adsurl = {https://ui.adsabs.harvard.edu/abs/2025A&A...694A.169S}
}

@ARTICLE{Strom_1973,
       author = {{Strom}, R.~G.},
        title = "{On the Relationship between Supernova Remnant Parameters and their Spectra}",
      journal = {Astronomy and Astrophysics},
         year = 1973,
        month = may,
       volume = {25},
        pages = {303-306},
       adsurl = {https://harvard.edu}
}

@ARTICLE{Taylor_2009,
       author = {{Taylor}, A.~R. and {Stil}, J.~M. and {Sunstrum}, C.},
        title = "{A Rotation Measure Image of the Sky}",
      journal = {\apj},
         year = 2009,
        month = sep,
       volume = {702},
       number = {2},
        pages = {1230-1236},
          doi = {10.1088/0004-637X/702/2/1230},
archivePrefix = {arXiv},
       eprint = {0907.1265},
 primaryClass = {astro-ph.CO},
       adsurl = {https://harvard.edu}
}

@article{Taylor_2024,
	doi = {10.1093/mnras/stae169},
	url = {https://doi.org},
	year = 2024,
	month = {jan},
	publisher = {Oxford University Press},
	volume = {528},
	number = {2},
	pages = {2511--2522},
	author = {A. R. Taylor and S. Sekhar and L. Heino and A. M. M. Scaife and J. Stil and M. Bowles and M. Jarvis and I. Heywood and J. D. Collier},
	title = {MIGHTEE polarization early science fields: the deep polarized sky},
	journal = {Monthly Notices of the Royal Astronomical Society}
}

@ARTICLE{Thomson_2023,
       author = {{Thomson}, Alec J.~M. and {McConnell}, David and {Lenc}, Emil and {Galvin}, Timothy J. and {Rudnick}, Lawrence and {Heald}, George and {Hale}, Catherine L. and {Duchesne}, Stefan W. and {Anderson}, Craig S. and {Carretti}, Ettore and {Federrath}, Christoph and {Gaensler}, B.~M. and {Harvey-Smith}, Lisa and {Haverkorn}, Marijke and {Hotan}, Aidan W. and {Ma}, Yik Ki and {Murphy}, Tara and {McClure-Griffiths}, N.~M. and {Moss}, Vanessa A. and {O'Sullivan}, Shane P. and {Raja}, Wasim and {Seta}, Amit and {Van Eck}, Cameron L. and {West}, Jennifer L. and {Whiting}, Matthew T. and {Wieringa}, Mark H.},
        title = "{The Rapid ASKAP Continuum Survey III: Spectra and Polarisation In Cutouts of Extragalactic Sources (SPICE-RACS) first data release}",
      journal = {\pasa},
         year = 2023,
        month = aug,
       volume = {40},
          eid = {e040},
        pages = {e040},
          doi = {10.1017/pasa.2023.38},
archivePrefix = {arXiv},
       eprint = {2307.07207},
 primaryClass = {astro-ph.GA},
       adsurl = {https://ui.adsabs.harvard.edu/abs/2023PASA...40...40T}
}

@misc{Thomson_2025,

doi = {10.25919/PBV3-WE20},

url = {https://data.csiro.au/collection/csiro:64891v7},

author = {Thomson, Alec and Galvin, Tim and Duchesne, Stefan and Lenc, Emil and Heald, George and Hlinka, Ondrej and Anderson, Craig and Osinga, Erik and Sebokolodi, Lerato and McClure-Griffiths, Naomi and Hutschenreuter, Sebastian and O'Sullivan, Shane and Akahori, Takuya and Gaensler, Bryan and Leahy, J Patrick and Ma, Jackie and Moss, Vanessa and Rudnick, Lawrence and Van Eck, Cameron and West, Jennifer},

title = {Spectra and Polarisation In Cutouts of Extragalactic sources from RACS Second Data Release (SPICE-RACS-DR2)},

publisher = {CSIRO},

year = {2025},

copyright = {Creative Commons Attribution 4.0 International}

}

@ARTICLE{Thomson_2026,
       author = {{Thomson}, Alec J.~M. and {Galvin}, Timothy J. and {Duchesne}, Stefan W. and {Lenc}, Emil and {Heald}, George and {Hlinka}, Ondrej and {Malik}, Sunil and {Anderson}, Craig S. and {Osinga}, Erik and {Baidoo}, Lerato and {McClure-Griffiths}, N.~M. and {Hutschenreuter}, Sebastian and {O'Sullivan}, Shane P. and {Akahori}, Takuya and {Gaensler}, B.~M. and {Leahy}, J.~P. and {Ma}, Y.~K. and {Moss}, Vanessa A. and {Rudnick}, L. and {Van Eck}, C.~L. and {West}, J.~L.},
        title = "{The Rapid ASKAP Continuum Survey VII: Spectra and Polarisation In Cutouts of Extragalactic Sources (SPICE-RACS) Second Data Release -- Unveiling the Magnetised Sky}",
      journal = {arXiv e-prints},
         year = 2026,
        month = may,
          eid = {arXiv:2605.16917},
        pages = {arXiv:2605.16917},
          doi = {10.48550/arXiv.2605.16917},
archivePrefix = {arXiv},
       eprint = {2605.16917},
 primaryClass = {astro-ph.GA},
       adsurl = {https://ui.adsabs.harvard.edu/abs/2026arXiv260516917T}
}

@misc{Van_Eck_2025,
       author = {{Van Eck}, Cameron L.},
        title = "{FRion: Time averaged correction of Faraday Rotation from the IONosphere}",
 howpublished = {Astrophysics Source Code Library, record ascl:2508.001},
         year = 2025,
        month = aug,
          eid = {ascl:2508.001},
archivePrefix = {ascl},
       eprint = {2508.001},
       adsurl = {https://ui.adsabs.harvard.edu/abs/2025ascl.soft08001V}
}

@article{Van_Eck_2026,
	doi = {10.3847/1538-4365/ae3dea},
	url = {https://doi.org},
	year = {2026},
	month = {mar},
	publisher = {The American Astronomical Society},
	volume = {283},
	number = {2},
	pages = {28},
	author = {Cameron L. Van Eck and Cormac R. Purcell and Lerato Baidoo and Alec J. M. Thomson and Yik Ki Ma and Lindsey Oberhelman and Erik Osinga and Shannon Vanderwoude and Jennifer L. West and Shinsuke Ideguchi and Dylan M. Par{\'e} and Jane F. Kaczmarek and Tony Willis and Takuya Akahori},
	title = {RM-Tools: Software for Analyzing Polarized Radio Spectra},
	journal = {The Astrophysical Journal Supplement Series},
	adsurl = {https://harvard.edu}
}

@ARTICLE{Wenger2000,
       author = {{Wenger}, M. and {Ochsenbein}, F. and {Egret}, D. and 
                 {Dubois}, P. and {Bonnarel}, F. and {Borde}, S. and 
                 {Genova}, F. and {Jaschek}, C. and {Lalo{\"e}}, S. and 
                 {Lesteven}, S. and {Monet}, A. and {Murtagh}, F. and 
                 {Ochsenbein}, F. and {Paturel}, G. and {Rousseau}, J.},
        title = "{The SIMBAD astronomical database. The CDS reference database for astronomical objects}",
      journal = {Astronomy and Astrophysics Supplement Series},
         year = 2000,
        volume = {143},
        pages = {9--22},
          doi = {10.1051/aas:2000332}
}

@ARTICLE{XuHan2014,
       author = {{Xu}, J. and {Han}, J.~L.},
        title = "{Extragalactic Rotation Measures and Magnetic Fields in Large-scale Structures}",
      journal = {Monthly Notices of the Royal Astronomical Society},
         year = 2014,
        volume = {442},
        pages = {3329--3340},
          doi = {10.1093/mnras/stu1059}
}
\bibliographystyle{aasjournalv7}

\end{document}